\documentclass[%
 aip,
 8pt,
 amsmath,amssymb,
reprint,%
]{revtex4-1}

\usepackage{lineno,hyperref}
\hypersetup{
    colorlinks=true,
    linkcolor=blue,
    filecolor=magenta,      
    urlcolor=blue,
}

\usepackage{amsmath}
\usepackage{amssymb}
\usepackage{graphicx}
\usepackage{dcolumn}
\usepackage{bm}
\usepackage{color}
\usepackage{multirow}
\usepackage{subfigure}
\usepackage{braket}
\usepackage[english]{babel}
\allowdisplaybreaks
\usepackage{cleveref}

\usepackage{array}
\newcolumntype{L}[1]{>{\raggedright\let\newline\\\arraybackslash\hspace{0pt}}m{#1}}
\newcolumntype{C}[1]{>{\centering\let\newline\\\arraybackslash\hspace{0pt}}m{#1}}
\newcolumntype{R}[1]{>{\raggedleft\let\newline\\\arraybackslash\hspace{0pt}}m{#1}}

\begin{document}


\author{Alessandro De Rosis}
\email{alessandro.derosis@manchester.ac.uk}
\affiliation{Department of Mechanical, Aerospace and Civil Engineering, The University of Manchester, Manchester M13 9PL, UK}

\author{Christophe Coreixas}
\email{christophe.coreixas@unige.ch}
\affiliation{Department of Computer Science, University of Geneva, 1204 Geneva, Switzerland}

\title{Multiphysics flow simulations using D3Q19 lattice Boltzmann methods based on central moments}


\begin{abstract}
In a recent work \href{https://aip.scitation.org/doi/abs/10.1063/1.5124719}{[A. De Rosis, R. Huang, and C. Coreixas, ``Universal formulation of central-moments-based lattice Boltzmann method with external forcing for the simulation of multiphysics phenomena", Phys. Fluids 31, 117102 (2019)]}, a multiple-relaxation-time lattice Boltzmann method (LBM) has been proposed by means of the D3Q27 discretization, where the collision stage is performed in the space of central moments (CMs). These quantities relax towards an elegant Galilean invariant equilibrium, and can also include the effect of external accelerations. Here, we investigate the possibility to adopt a coarser lattice composed of 19 discrete velocities only. The consequences of such a choice are evaluated in terms of accuracy and stability through multiphysics benchmark problems based on single-, multi-phase and magnetohydrodynamics flow simulations. In the end, it is shown that the reduction from 27 to 19 discrete velocities have only little impact on the accuracy and stability of the CM-LBM for moderate Reynolds number flows in the weakly compressible regime.
\end{abstract}


\maketitle

\section{Motivation}
Numerical simulations of viscous fluid flows are routinely performed by scientists involved in both the academic and industrial sectors. These simulations can be approached by different viewpoints. The most common one is the so-called \textit{macroscopic} or continuum-based level, with the problem being governed by the (in)compressible Navier-Stokes equations (NSEs). As a drastic alternative, molecular dynamics (MD) idealizes a certain volume of fluid by a finite set of particles obeying Newton's law. Interestingly, MD possesses the great advantage to introduce physics properties at the \textit{microscopic} level. However, if one wants to simulate a relatively large-scale realistic fluid problem, the number of required particles becomes very high, thus dramatically increasing the computational cost. Between MD and NSEs there is another level, called \textit{mesoscopic} or kinetic, and its practical implementation is represented by the lattice Boltzmann method (LBM)~\cite{benzi1992lattice, KRUGER_Book_2017, succi2018lattice}.\\
\indent Roughly speaking, the LBM recovers the behavior of fluid dynamics from the motion of populations (or distribution functions) of fictitious particles, that collide and stream along the links of a fixed Cartesian lattice, representing the fluid domain. The collision stage is the core of any LB algorithm because it retains the whole flow physics. The so-called BGK collision operator (named after its authors Bhatnagar Gross and Krook~\cite{bhatnagar1954model}) represents the simplest, yet effective, and most popular approximation, where all the populations relax with the same common 
rate towards a discrete equilibrium state. The latter is usually derived by applying a Gauss-Hermite quadrature to the continuous Maxwellian distribution~\cite{SHAN_PRL_80_1998,shan2006kinetic}. Despite its simplicity and popularity, the BGK-LBM is unsuitable for the prediction of turbulent flows as it becomes rapidly unstable in the limit of vanishing viscosity. This is mainly (though not exclusively) due to the intrinsic unavoidable presence of non-hydrodynamic ghost modes, which undermine the stability of the algorithm through spurious couplings with hydrodynamic ones~\cite{DELLAR_PRE_65_2002,DELLAR_PA_362_2006,ADHIKARI_PRE_78_2008, WISSOCQ_JCP_380_2019}. Originally proposed to introduce free parameters that could be used to either increase the stability (by damping non-hydrodynamic modes) and/or the validity range of LBMs (variable Prandtl number, etc), the multiple-relaxation-time (MRT) LBM introduced the idea to perform the collision in a space of raw (or absolute) moments of different order
~\cite{DHUMIERE_PAA_159_1992,LALLEMAND_PRE_61_2000,d2002multiple}. While second-order ones entail the flow and relax with a certain frequency directly linked to the fluid kinematic and bulk viscosities, higher-order ones address the above-mentioned ghost modes and the corresponding frequencies can be (almost) freely tuned to improve the stability of the resulting LBM
~\cite{HOSSEINI_PRE_99_2019b,MASSET_JFM_897_2020,WISSOCQ_ARXIV_2020_07353}.\\
\indent Another problem affecting the LBM is the lack of Galilean invariance at \emph{all} the orders, which results in velocity-dependent transport coefficients~\cite{NIE_EPL_81_2008,DELLAR_JCP_259_2014,SHAN_PRE_100_2019}. For standard lattices, this is due to the fact that populations (for the BGK) and moments (for the MRT) relax to equilibrium states derived through a second-order truncated Taylor expansion in the local Mach number of the above-mentioned continuous Maxwellian distribution~\cite{DELLAR_JCP_259_2014}. 
In 2006, Geier \textit{et al.}~\cite{geier2006cascaded} argued that this problem could be solved by performing the collision stage in the space of central moments (CMs), obtained by shifting the lattice directions by the local fluid velocity. Moreover, they imposed the match between the CMs of the continuous Maxwellian distribution and those of the discrete counterpart. Consequently, this methodology assumes that equilibrium CMs are unchanged by the velocity discretization of the Boltzmann equation, which might not be true depending on both equilibrium CMs of interest and on the considered lattice of discrete velocities. The resulting scheme, called cascaded LBM, remarkably outperformed both the BGK- and MRT-LBMs in terms of stability thanks to: (1) the implicit use of an extended equilibrium, and (2) the equilibration of CMs related to high-order moments and bulk viscosity~\cite{COREIXAS_RSTA_378_2020}.\\
\indent More recently, it was demonstrated that it is possible to adopt a CMs-based procedure where moments relax to whatever discrete equilibrium state and for whatever lattice discretization~\cite{derosis2016epl_d2q9, de2017nonorthogonal, de2017central_shallow, de2018advanced, doi:10.1021/acs.jpcb.9b10989}. It is possible to switch from populations to CMs (and vice versa) by simply multiplying (or dividing) by a transformation matrix that depends on the adopted lattice and the local fluid velocity. These early attempts relied on a second-order equilibrium state in the D2Q9 and D3Q27 lattice velocity spaces. The adoption of such a simple (yet naive and incomplete) equilibrium distribution generates a non-negligible number of non-zero velocity-dependent equilibrium CMs. As a consequence, post-collision populations are functions of all these terms, which leads to a huge computational overhead~\cite{fei2018three}. However, by including the correct higher-order terms in the definition of the discrete equilibrium~\cite{malaspinas2015increasing,coreixas2017recursive, COREIXAS_PhD_2018,COREIXAS_PRE_100_2019}, the methodology outlined in Ref.~\cite{de2017nonorthogonal} was proven to lead to Galilean invariant CMs with the D3Q27 discretization~\cite{derosisHermite}. This D3Q27-CM-LBM was further shown to recover the behavior of the original cascaded LBM by only relying on the correct set of Hermite polynomials. Since this family of polynomials is tightly linked to the velocity discretization of the Boltzmann equation~\cite{grad1949kinetic,SHAN_PRL_80_1998,shan2006kinetic}, the correct set of Hermite polynomials is known in an \emph{a priori} way, hence ensuring its consistency for any kind of lattices. The derivation of D3Q19-LBMs based on Galilean invariants CMs makes no exception, and only requires to apply simple pruning rules on its D3Q27 counterpart~\cite{COREIXAS_PRE_100_2019}.\\
\indent During the past three decades, numerous velocity discretizations have been proposed for the simulation of isothermal and weakly compressible flows. For example, lattices based on 15 and 19 discrete velocities were introduced by Qian \textit{et al.}~\cite{QIAN_EPL_17_1992}, while Ladd~\cite{ladd1994numerical} discarded the velocity corresponding to particles at rest, hence, ending up with a D3Q18 lattice. These three velocity discretizations directly flow from the more general D3Q27 lattice through prunning~\cite{KRUGER_Book_2017}. This is one of the common way to reduce the number of discrete velocities, while keeping most of the macroscopic properties intact for LBMs based on second-order equilibria. The other popular strategy is the moment-matching approach which leads, e.g., to the smallest lattice for the simulation of fluid flows, namely, the D3Q13 lattice~\cite{DHUMIERES_PRE_63_2001}.\\
\indent The D3Q19 has been the default choice for a long time, notably due to checkerboard instabilities arising for smaller lattices~\cite{kandhai1999implementation,d2002multiple,KRUGER_Book_2017}. 
There has been sporadic interest using the D3Q27 lattice for problems where the rotational invariance of the numerical solution is of particular concern
~\cite{mayer2006direct, white2011rotational, geller2013turbulent, kang2013effect, silva2014truncation, augier2014rotational}. However, it was recently proven that numerical errors --notably those induced by the equilibrium-- are at the origin of the latter rotational problem instead of the lattice itself~\cite{BAUER_JCP_405_2020}. This is in agreement with the fact that several commercial solvers (e.g., PowerFLOW and ProLB) rely on D3Q19 formulations --with regularized/filtered collision models whose equilibria include high-order velocity terms-- without suffering from such anistropic issues~\cite{LEW_AIAA_2755_2014,KHORRAMI_CEAS_AERO_2019,CHEN_PS_Accepted_2019,SENGISSEN_AIAA_2993_2015,FENG_ESSOA_Submitted_2020}.\\
\indent Surprisingly, most research works rely on  D3Q27-CM-LBMs. This is also the case of XFlow software which is a CM-LB solver dedicated to industry-oriented flow simulations~\cite{holman2012solution,CHAVEZ_Energies_13_2020}. Nevertheless, it is possible to find in the literature formulations of CM-LBMs based on D3Q15 and D3Q19 lattices, even though the latter are pretty rare. As an example, Premnath \& Banerjee~\cite{premnath2011three} showed a comparison between the D3Q27 and D3Q15 lattices using these techniques, and they obtained similar results for low Reynolds and Mach number flow simulations. More recently, Fei \textit{et al.}~\cite{fei2018three} have presented an improved implementation of the cascaded scheme for both D3Q27 and D3Q19 lattices, even though, no comparative study was performed by the authors. In the end, it is still not clear if one can reduce the lattice size in the context of CM-LBMs without deteriorating the accuracy of the solver. This is even more true in the context of multiphysics flow simulations, for which, to the best of the authors' knowledge, no systematic accuracy/stability comparison study can be found in the literature. \\ 
\indent In this paper, we then aim at deriving a D3Q19-CM-LBM for the simulation of multiphysics flows with or without external acceleration. To make sure this is not done at the expense of accuracy and/or stability, both points will be at the center of the comparative study that will be carried out throughout the paper. 
The rest of the paper is organized as follows. Sec.~\ref{SEC:II} presents a detailed derivation of our D3Q19-CM-LBM which is based on extended formualtions of the equilibrium and forcing terms. This approach is thoroughly tested against several well-defined and consolidated benchmark problems in Sec.~\ref{SEC:III}. Some conclusions are drawn in Sec.~\ref{SEC:IV}. Finally, additional details are given in Appendices: (\ref{app:AppExtendedEquilibrium}) impact of extended equilibria on stability and accuracy, (\ref{app:AppForcingQ19}) derivation of the extended forcing term, (\ref{app:AppRMQ19}) raw moment formulation of our approach, (\ref{app:Q27}) recalls on the D3Q27-CM formulation, and (\ref{app:AppCG}) color-gradient method algorithm.

\section{Methodology}
\label{SEC:II}
In this Section, we first recall the classic D3Q19-BGK-LBM. Then, we derive a collision operator in the space of CMs and present the D3Q19-CM-LBM. Finally, we demonstrate that the classical D3Q19-MRT-LBM based on the relaxation of raw (or absolute) moments can be interpreted as a particular case of our D3Q19-CM-LBM. If not otherwise stated, the LB unit system will be used henceforth, {\color{black}where the grid spacing and the time step are both equal to 1}.
\subsection{D3Q19-BGK-LBM}
Let us consider an Eulerian basis $\bm{x}=[x,y,z]$. The lattice Boltzmann equation (LBE) 
\begin{equation}
f_i(\bm{x}+\bm{c}_i, t+1) = f_i(\bm{x}, t) + \Omega_i(\bm{x}, t) + \left(1-\frac{\omega}{2} \right)F_i(\bm{x}, t),
\end{equation}
predicts the space and time evolution of the particle distribution functions $\displaystyle | f_{i}\rangle = \left[ f_0,\, \ldots ,\, f_i,\, \ldots,\, f_{18}    \right]^{\top}$ that collide and stream along the generic link $i=0 \ldots 18$ with the discrete velocities $\displaystyle \bm{c}_i=[| c_{ix}\rangle ,\, | c_{iy}\rangle ,\, | c_{iz}\rangle]$ defined as
\begin{widetext}
\begin{align}
| c_{ix}\rangle &= [0, -1, \phantom{-}0, \phantom{-}0, -1, -1, -1, -1, \phantom{-}0, \phantom{-}0, \phantom{-}1, \phantom{-}0, \phantom{-}0, \phantom{-}1, \phantom{-}1, \phantom{-}1, \phantom{-}1, \phantom{-}0, \phantom{-}0]^{\top}, \nonumber\\
| c_{iy}\rangle &= [0, \phantom{-}0, -1, \phantom{-}0, -1, \phantom{-}1, \phantom{-}0, \phantom{-}0, -1, -1, \phantom{-}0, \phantom{-}1, \phantom{-}0, \phantom{-}1, -1, \phantom{-}0, \phantom{-}0, \phantom{-}1, \phantom{-}1]^{\top}, \nonumber\\
| c_{iz}\rangle &= [0, \phantom{-}0, \phantom{-}0, -1, \phantom{-}0, \phantom{-}0, -1, \phantom{-}1, -1, \phantom{-}1, \phantom{-}0, \phantom{-}0, \phantom{-}1, \phantom{-}0, \phantom{-}0, \phantom{-}1, -1, \phantom{-}1, -1]^{\top}.
\end{align}
\end{widetext}
As usual, this numerical scheme can be divided into two parts, i.e., collision:
\begin{equation}
f_i^{\star}(\bm{x}, t) = f_i(\bm{x}, t) + \Omega_i(\bm{x}, t) + \left(1-\frac{\omega}{2} \right)F_i(\bm{x}, t),
\end{equation}
and streaming:
\begin{equation}\label{streaming}
f_i(\bm{x}+\bm{c}_i, t+1) = f_i^{\star}(\bm{x}, t),
\end{equation}
where the superscript $\star$ denotes post-collision quantities here and henceforth. Let us implicitly assume the dependence on the spatial position $\bm{x}$ and the time $t$ in the following. Within the BGK approximation, the collision operator $\Omega_i$ can be written as a relaxation of the populations towards an equilibrium state $f_i^{eq}$, i.e.
\begin{equation} \label{operatorBGK}
\Omega_i = \omega  \left( f_i^{eq} - f_i \right),
\end{equation}
where $\omega$ is a relaxation frequency that is linked to the fluid kinematic viscosity $\nu$ as 
\begin{equation}
\nu = \left( \frac{1}{\omega}-\frac{1}{2} \right)c_s^2,
\end{equation}
$c_s=1/\sqrt{3}$ being the lattice sound speed of the D3Q19 velocity discretization~\cite{KRUGER_Book_2017, succi2018lattice}. The source term $F_i$ is usually treated according to the popular model by Guo et al.~\citep{GUO_PRE_65_2002}. The choice of the equilibrium populations is instrumental to recover the correct physics of a phenomenon. At a first glance, one might be tempted to use the popular second-order truncated expression~\cite{QIAN_EPL_17_1992}
\begin{equation}\label{Eq:2ndEquilibrium}
f_i^{eq} = w_i \rho \left[  1+ \frac{\bm{c}_i \cdot  \bm{u}}{c_s^2}+\frac{\left(  \bm{c}_i \cdot  \bm{u} \right)^2}{2 c_s^4}-\frac{ \bm{u}^2}{2  c_s^2} \right],
\end{equation}
$\rho$ and $\bm{u} = [u_x,u_y,u_z]$ being the mass density and the flow velocity, respectively, and the weights are
\begin{widetext}
\begin{equation}
w = \left[w_0 ,\, w_s,\, w_s,\, w_s,\, w_l,\, w_l,\, w_l,\, w_l,\, w_l,\, w_l ,\,w_s,\, w_s,\, w_s,\, w_l ,\,w_l ,\,w_l,\, w_l ,\,w_l,\, w_l \right],
\end{equation}
\end{widetext}
with $w_0=1/3$, $w_s=1/18$ and $w_l=1/36$. Lattice directions $ \bm{c}_i$ and weights $w_i$ are defined according to Ref.~\cite{latt2007technical}, in order to further reduce memory consumption thanks to the ``swap trick''. 
However, several authors demonstrated that the full potential of any LB discretization (in terms of physical and numerical properties) can only be achieved by using the complete allowable set of Hermite polynomials~\cite{ADHIKARI_PRE_78_2008,malaspinas2015increasing,coreixas2017recursive,COREIXAS_PRE_100_2019,COREIXAS_RSTA_378_2020}. This is true for all lattices derived through a tensor-product of lower-order ones (e.g., D2Q9 and D3Q27), but with the D3Q19 lattice a pruning strategy must be adopted. Using the latter strategy, Coreixas et al. proposed a derivation of the equilibrium that is compliant with all collision models (Appendix H of Ref.~\cite{COREIXAS_PRE_100_2019}), and the corresponding expressions are
\begin{widetext}
\begin{align}
f^{eq}_{0}&=\tfrac{\rho}{3} \big[1 -(u_x^2+u_y^2+u_z^2)+ 3(u_x^2 u_y^2+u_x^2 u_z^2+u_y^2 u_z^2)\big],\nonumber\\[0.1cm]
f^{eq}_{1}&= \tfrac{\rho}{18} \big[1 -3 u_x + 3 (\phantom{-}u_x^2- u_y^2- u_z^2) + 9  (u_x u_y^2 + u_x u_z^2)-9 (u_x^2 u_y^2 + u_x^2 u_z^2)\big],\nonumber\\[0.1cm]
f^{eq}_{2}&= \tfrac{\rho}{18} \big[1 -3  u_y +3 (- u_x^2+  u_y^2- u_z^2) +9  (u_x^2 u_y + u_y u_z^2)-9 (u_x^2 u_y^2+ u_y^2 u_z^2)\big],\nonumber\\[0.1cm]
f^{eq}_{3}&= \tfrac{\rho}{18} \big[1 -3  u_z +3(- u_x^2- u_y^2+  u_z^2) +9  (u_x^2 u_z + u_y^2 u_z)-9 (u_x^2 u_z^2+ u_y^2 u_z^2)\big],\nonumber \\[0.1cm]
f^{eq}_{4} &= \tfrac{\rho}{36}  \big[1-3(\phantom{-} u_x +  u_y) + 3 (u_x^2+ u_y^2)+9  u_x u_y - 9(\phantom{-}u_x^2 u_y+ u_x u_y^2)+9 u_x^2 u_y^2\big],\nonumber\\[0.1cm]
f^{eq}_{5} &= \tfrac{\rho}{36}  \big[1+3(- u_x +  u_y) + 3 (u_x^2+ u_y^2)-9  u_x u_y + 9(\phantom{-}u_x^2 u_y- u_x u_y^2)+9 u_x^2 u_y^2\big],\nonumber\\[0.1cm]
f^{eq}_{6} &= \tfrac{\rho}{36}  \big[1-3(\phantom{-} u_x +  u_z) + 3 (u_x^2+ u_z^2)+9  u_x u_z - 9(\phantom{-}u_x^2 u_z+ u_x u_z^2)+9 u_x^2 u_z^2\big],\nonumber\\[0.1cm]
f^{eq}_{7} &= \tfrac{\rho}{36}  \big[1+3(- u_x +  u_z) + 3 (u_x^2+ u_z^2)-9  u_x u_z + 9(\phantom{-}u_x^2 u_z- u_x u_z^2)+9 u_x^2 u_z^2\big],\nonumber\\[0.1cm]
f^{eq}_{8} &= \tfrac{\rho}{36}  \big[1-3(\phantom{-} u_y +  u_z) + 3 (u_y^2+ u_z^2)+9  u_y u_z - 9(\phantom{-}u_y^2 u_z+ u_y u_z^2)+9 u_y^2 u_z^2\big],\nonumber\\[0.1cm]
f^{eq}_{9} &= \tfrac{\rho}{36}  \big[1+3(- u_y +  u_z) + 3 (u_y^2+ u_z^2)-9  u_y u_z + 9(\phantom{-}u_y^2 u_z- u_y u_z^2)+9 u_y^2 u_z^2\big],\label{eq:EqQ19RM}\\[0.1cm]
f^{eq}_{10}&= \tfrac{\rho}{18} \big[1 +3 u_x + 3 (\phantom{-}u_x^2- u_y^2- u_z^2) - 9  (u_x u_y^2 + u_x u_z^2)-9 (u_x^2 u_y^2 + u_x^2 u_z^2)\big],\nonumber\\[0.1cm]
f^{eq}_{11}&= \tfrac{\rho}{18} \big[1 +3  u_y +3 (- u_x^2+  u_y^2- u_z^2) -9  (u_x^2 u_y + u_y u_z^2)-9 (u_x^2 u_y^2+ u_y^2 u_z^2)\big],\nonumber\\[0.1cm]
f^{eq}_{12}&= \tfrac{\rho}{18} \big[1 +3  u_z +3(- u_x^2- u_y^2+  u_z^2) -9  (u_x^2 u_z + u_y^2 u_z)-9 (u_x^2 u_z^2+ u_y^2 u_z^2)\big],\nonumber \\[0.1cm]
f^{eq}_{13} &= \tfrac{\rho}{36} \big[1+3(\phantom{-} u_x +  u_y) + 3 (u_x^2+ u_y^2)+9  u_x u_y + 9(\phantom{-}u_x^2 u_y+ u_x u_y^2)+9 u_x^2 u_y^2\big],\nonumber\\[0.1cm]
f^{eq}_{14} &= \tfrac{\rho}{36} \big[1+3(\phantom{-} u_x -  u_y) + 3 (u_x^2+ u_y^2)-9  u_x u_y + 9( -u_x^2 u_y+ u_x u_y^2)+9 u_x^2 u_y^2\big],\nonumber\\[0.1cm]
f^{eq}_{15} &= \tfrac{\rho}{36} \big[1+3(\phantom{-} u_x +  u_z) + 3 (u_x^2+ u_z^2)+9  u_x u_z + 9(\phantom{-}u_x^2 u_z+ u_x u_z^2)+9 u_x^2 u_z^2\big],\nonumber\\[0.1cm]
f^{eq}_{16} &= \tfrac{\rho}{36} \big[1+3(\phantom{-} u_x -  u_z) + 3 (u_x^2+ u_z^2)-9  u_x u_z + 9( -u_x^2 u_z+ u_x u_z^2)+9 u_x^2 u_z^2\big],\nonumber\\[0.1cm]
f^{eq}_{17} &= \tfrac{\rho}{36} \big[1+3(\phantom{-} u_y +  u_z) + 3 (u_y^2+ u_z^2)+9  u_y u_z + 9(\phantom{-}u_y^2 u_z+ u_y u_z^2)+9 u_y^2 u_z^2\big],\nonumber\\[0.1cm]
f^{eq}_{18} &= \tfrac{\rho}{36} \big[1+3(\phantom{-} u_y -  u_z) + 3 (u_y^2+ u_z^2)-9  u_y u_z + 9( -u_y^2 u_z+ u_y u_z^2)+9 u_y^2 u_z^2\big].\nonumber
\end{align}
\end{widetext}
As usual, macroscopic variables are computed as the zeroth- and first-order moments of the populations:
\begin{eqnarray}
\rho &=& \sum_i f_i , \nonumber \\
\rho \bm{u} &=& \sum_i f_i \bm{c}_i.\label{macro}
\end{eqnarray}
It should be noted that the $u^3$ terms restore the Galilean invariance for shear flows aligned  with the coordinate axes~\cite{hazi2006cubic}. However, a complete restoration is impossible for standard lattices because one cannot add the diagonal terms $u_x^3$, $u_y^3$ and $u_z^3$ to the components of the third moment of $f_i^{eq}$. The partial restoration of Galilean invariance leads to an anisotropic stress-strain relation that can increase the errors for shear flows inclined to axes~\cite{dellar2014lattice}, that can only be removed through correction terms~\cite{FENG_JCP_394_2019,HOSSEINI_RSTA_378_2020,RENARD_ARXIV_2020_03644,RENARD_ARXIV_2020_08477}. Nevertheless, the extended equilibrium~(\ref{eq:EqQ19RM}) should be preferred to its second-order counter part~(\ref{Eq:2ndEquilibrium}), as it allows for better stability for simulations at moderate Mach numbers, and in the low viscosity regime (see App.~\ref{app:AppExtendedEquilibrium} for more details).

\subsection{General D3Q19-MRT-LBM}
The lattice Boltzmann equation (LBE) with the forcing term can be generally expressed as~\citep{McCracken2005,Fei2017}
\begin{widetext}
\begin{equation}\label{lbe}
\ket{f_i(\boldsymbol{x}+\boldsymbol{c}_i, t+1)} 
= \ket{f_i(\boldsymbol{x}, t)} 
+ {\bm \Lambda} [ \ket{f_i^{\mathrm{eq}}(\boldsymbol{x}, t)} 
-\ket{f_i(\boldsymbol{x}, t)} ] 
+ (\mathbf{I} - {\bm \Lambda}/2 ) \ket{F_i(\boldsymbol{x}, t)},
\end{equation}
\end{widetext}
{\color{black} where $\ket{\bullet}$ denotes a row vector.}Notice that Eq.~(\ref{lbe}) collapses into the aforementioned BGK-LBM if the collision matrix is set to ${\bm \Lambda} = \omega \mathbf{I}$, where $\mathbf{I}$ is the unit tensor. The term $F_i$ accounts for external body forces $\boldsymbol{F} = [F_x, \,F_y, \, F_z]$ and its role will be elucidated later. Its prefactor accounts for discrete effects originating from the change of variables that aims at obtaining a numerical scheme explicit in time~\cite{GUO_PRE_65_2002}. Again, the LBE can be divided into two steps, i.e., collision
\begin{widetext}
\begin{equation}\label{collision}
\ket{f_i^{\star}(\boldsymbol{x}, t)} 
= \ket{f_i(\boldsymbol{x}, t)} 
+ {\bm \Lambda} [ \ket{f_i^{\mathrm{eq}}(\boldsymbol{x}, t)} -\ket{f_i(\boldsymbol{x}, t)} ] + (\mathbf{I}- {\bm \Lambda}/2 ) \ket{F_i(\boldsymbol{x}, t)},
\end{equation}
\end{widetext}
and streaming
\begin{equation}\label{streaming_f}
\ket{f_i(\boldsymbol{x}+\boldsymbol{c}_i, t+1)}  = \ket{f_i^{\star}(\boldsymbol{x}, t)} .
\end{equation}

Let us first neglect the presence of $F_i$. In order to build a CMs-based collision operator, the lattice directions are shifted by the local fluid velocity~\citep{geier2006cascaded}. These shifted or peculiar discrete velocities $\displaystyle \bar{\bm{c}}_i=[\bra{\bar{c}_{ix}} ,\, \bra{\bar{c}_{iy}} ,\, \bra{\bar{c}_{iz}}]$ are defined as
\begin{eqnarray}
\bra{\bar{c}_{ix}} &=&  \bra{c_{ix}- u_x}, \nonumber \\
\bra{\bar{c}_{iy}} &=& \bra{c_{iy}- u_y}, \nonumber \\
\bra{\bar{c}_{iz}} &=& \bra{c_{iz}- u_z},
\end{eqnarray}
{\color{black} where $\bra{\bullet}$ denotes a column vector.} In order to apply the collision step in the CM space, one must choose a suitable basis. In the present case, the following transformation matrix (from populations to CMs) is proposed:
\begin{equation} \label{eq:matrix}
{\mathbf T} = 
\left [
\begin{array}{c}
\bra{|\boldsymbol{c}_i|^0} \\
\bra{ \bar{c}_{ix}}\\ 
\bra{ \bar{c}_{iy}} \\
\bra{ \bar{c}_{iz}} \\  
\bra{ \bar{c}_{ix}^2+ \bar{c}_{iy}^2 + \bar{c}_{iz}^2 } \\
\bra{ \bar{c}_{ix}^2- \bar{c}_{iy}^2 } \\ 
\bra{ \bar{c}_{iy}^2- \bar{c}_{iz}^2}\\
\bra{ \bar{c}_{ix} \bar{c}_{iy}} \\
\bra{ \bar{c}_{ix} \bar{c}_{iz} } \\ 
\bra{ \bar{c}_{iy} \bar{c}_{iz}} \\
\bra{\bar{c}_{ix}^2\bar{c}_{iy}}\\
\bra{\bar{c}_{ix}\bar{c}_{iy}^2} \\
\bra{ \bar{c}_{ix}^2\bar{c}_{iz}}\\
\bra{ \bar{c}_{ix}\bar{c}_{iz}^2} \\
\bra{\bar{c}_{iy}^2\bar{c}_{iz}} \\
\bra{ \bar{c}_{iy}\bar{c}_{iz}^2} \\
\bra{ \bar{c}_{ix}^2\bar{c}_{iy}^2} \\
\bra{ \bar{c}_{ix}^2\bar{c}_{iz}^2} \\
\bra{\bar{c}_{iy}^2\bar{c}_{iy}^2}
\end{array}
\right] .
\end{equation}	
This basis directly flows from its D3Q27 counterpart where monomials related to discrete velocities $(\pm1,\pm1,\pm1)$ are discarded~\cite{FEI_PRE_97_2018,COREIXAS_PRE_100_2019}. In addition, by decoupling moments related to compression/dilation phenomena (trace of the second-order-moment tensor) from those controlling shear phenomena (off-diagonal terms), it is possible to adjust the bulk viscosity independently from its shear counterpart thanks to a \emph{diagonal} relaxation matrix~\cite{KRUGER_Book_2017,FEI_PRE_97_2018,COREIXAS_PRE_100_2019}. The relaxation matrix in the populations space then reads ${\bm \Lambda} = \mathbf{T}^{-1} \mathbf{K} \mathbf{T}$, where in the present case $\mathbf{K} = \mathrm{diag}[1,1,1,1,1,\omega,\omega,\omega,\omega,\omega,1,\ldots,1]$ is the $19 \times 19$ relaxation matrix in the CMs space. The latter has been chosen in order to (1) take into account external forces (if needed be) through the non-zero first four relaxation frequencies, (2) impose the kinematic viscosity $\nu$ thanks to $\omega=1/(\nu/c_s^2 +1/2)$, and (3) improve the numerical stability via the equilibration of bulk and high-order CMs.

 Let us collect pre-collision, equilibrium and post-collision CMs as
\begin{eqnarray}
\ket{k_i} &=& \left[ k_0,\, \ldots, \, k_i,\, \ldots, \, k_{18}   \right]^\top, \nonumber \\
\ket{k_i^{\mathrm{eq}}} &=& \left[ k_0^{\mathrm{eq}},\, \ldots, \, k_i^{\mathrm{eq}},\, \ldots, \, k_{18}^{\mathrm{eq}}   \right]^\top, \nonumber \\
\ket{k_i^{\star}} &=& \left[ k_0^{\star},\, \ldots, \, k_i^{\star},\, \ldots, \, k_{18}^{\star}   \right]^\top.
\end{eqnarray}
respectively. 
The first two quantities are evaluated by applying the matrix ${\mathbf{T}}$ to the corresponding distribution, that is
\begin{eqnarray}
\ket{k_i} 					  &=&  {\mathbf{T}} \ket{f_i}, \nonumber \\
\ket{k_i^{\mathrm{eq}}}  &=&  {\mathbf{T}} \ket{f_i^{\mathrm{eq}}} ,
\end{eqnarray}
Interestingly, applying the transformation matrix $\mathbf{T}$ to equilibrium populations in Eqs.~(\ref{eq:EqQ19RM}) generates the following equilibrium CMs:
\begin{eqnarray}
k_0^{eq} &= \rho, \nonumber \\ 
k_4^{eq} &= 3\rho c_s^2, \nonumber \\ 
k_{16}^{eq} &= \rho c_s^4, \nonumber \\ 
k_{17}^{eq} &= \rho c_s^4,  \nonumber \\ 
k_{18}^{eq} &= \rho c_s^4,
\end{eqnarray}
while the remaining terms are equal to zero. It is of interest to notice that the equilibrium CMs are Galilean invariant, as no dependence on the fluid velocity is present. This is consistent with the theoretical findings in Ref.~\cite{derosisHermite}, where it has been demonstrated that, for tensor-product-based lattices (D2Q9 and D3Q27), Galilean invariant equilibrium CMs are found if the transformation matrix $\mathbf{T}$ is applied to discrete equilibrium populations accounting for the correct high-order Hermite polynomials -- those based on tensor products of second-order Hermite polynomials. It is worth noting that one could further discard the remaining lattice-dependent terms (those proportional to the lattice constant $c_s$) thanks to the central-Hermite moment approach, as explained in Ref.~\cite{COREIXAS_PRE_100_2019}. The collision process takes place as
\begin{equation}\label{eq:posColl}
\ket{k_i^{\star}} 
= \left( \mathbf{I}- \mathbf{K} \right) \mathbf{T} \ket{f_i} 
+ \mathbf{K}  \mathbf{T} \ket{f_i^{\mathrm{eq}}}
= \left( \mathbf{I}- \mathbf{K} \right)  \ket{k_i} 
+ \mathbf{K}   \ket{k_i^{\mathrm{eq}}} .
\end{equation}
After the collision, non-zero CMs read as follows:
\begin{eqnarray}
k_0^{\star} &=& \rho, \nonumber \\
k_4^{\star} &=& 3\rho c_s^2, \nonumber \\
k_5^{\star} &=& \left(1-\omega \right)k_5, \nonumber \\
k_6^{\star} &=& \left(1-\omega \right)k_6 \nonumber \\
k_7^{\star} &=& \left(1-\omega \right)k_7, \nonumber \\
k_8^{\star} &=& \left(1-\omega \right)k_8, \nonumber \\
k_9^{\star} &=& \left(1-\omega \right)k_9, \nonumber \\
k_{16}^{\star} &=& \rho c_s^4, \nonumber \\
k_{17}^{\star} &=& \rho c_s^4, \nonumber \\
k_{18}^{\star} &=& \rho c_s^4,  \label{post_collision_central_moments}
\end{eqnarray}
where pre-collision CMs are
\begin{eqnarray}
k_5 &=& \sum_i f_i (\bar{c}_{ix}^2- \bar{c}_{iy}^2), \nonumber \\
k_6 &=& \sum_i f_i (\bar{c}_{iy}^2- \bar{c}_{iz}^2), \nonumber \\
k_7 &=& \sum_i f_i \bar{c}_{ix} \bar{c}_{iy}, \nonumber \\
k_8 &=& \sum_i f_i \bar{c}_{ix} \bar{c}_{iz}, \nonumber \\
k_9 &=& \sum_i f_i \bar{c}_{iy} \bar{c}_{iz}. \label{pre_collision_central_moments}
\end{eqnarray}
Now, we are in the position to reconstruct post-collision populations
\begin{equation}\label{system}
\ket{f_i^{\star}} 
= \mathbf{T}^{-1} \ket{k_i^{\star} },
\end{equation}
with $|f_i^{\star} \rangle = [f_0^{\star} ,\, \ldots f_i^{\star}, \, \ldots f_{18}^{\star} ]^{\top}$. Eventually, populations are streamed (see Eq.~(\ref{streaming})) and macroscopic variables are computed by Eq.~(\ref{macro}).

In the present model, the forcing term $F_i$ is accounted for through the collision step in Eq.~(\ref{lbe}). The latter is applied in the CM space, and consequently, it requires the computation of the forcing term CMs, $R_i$, following
\begin{equation}
\ket{R_i} = \mathbf{T} \ket{F_i }.
\end{equation}
In presence of external forces, post-collision CMs then read as
\begin{widetext}
\begin{equation}\label{eq:posColl0}
\ket{k_i^{\star}} 
= \left( \mathbf{I}- \mathbf{K} \right) \mathbf{T} \ket{f_i} 
+ \mathbf{K}  \mathbf{T} \ket{f_i^{\mathrm{eq}}} + \left( \mathbf{I}
- \frac{\mathbf{K}}{2} \right) \mathbf{T} \ket{F_i}  
= \left( \mathbf{I}- \mathbf{K} \right)  \ket{k_i} 
+ \mathbf{K}   \ket{k_i^{\mathrm{eq}}} + \left( \mathbf{I}
- \frac{\mathbf{K}}{2} \right)  \ket{R_i},
\end{equation}
\end{widetext}
with
\begin{eqnarray}
k_0^{\star} &=& \rho, \nonumber \\
k_1^{\star} &=& F_x/2, \nonumber \\
k_2^{\star} &=& F_y/2, \nonumber \\
k_3^{\star} &=& F_z/2, \nonumber \\
k_4^{\star} &=& 3 \rho c_s^2, \nonumber \\
k_5^{\star} &=& \left(1-\omega \right)k_5, \nonumber \\
k_6^{\star} &=& \left(1-\omega \right)k_6 \nonumber \\
k_7^{\star} &=& \left(1-\omega \right)k_7, \nonumber \\
k_8^{\star} &=& \left(1-\omega \right)k_8, \nonumber \\
k_9^{\star} &=& \left(1-\omega \right)k_9, \nonumber \\
k_{10}^{\star} &=& F_y c_s^2 /2, \nonumber \\
k_{11}^{\star} &=& F_x c_s^2 /2, \nonumber \\
k_{12}^{\star} &=& F_z c_s^2 /2, \nonumber \\
k_{13}^{\star} &=& F_x c_s^2 /2, \nonumber \\
k_{14}^{\star} &=& F_z c_s^2 /2, \nonumber \\
k_{15}^{\star} &=& F_y c_s^2 /2, \nonumber \\
k_{16}^{\star} &=& \rho c_s^4, \nonumber \\
k_{17}^{\star} &=& \rho c_s^4, \nonumber \\
k_{18}^{\star} &=& \rho c_s^4,
\end{eqnarray}
if one assumes that CMs of the forcing term are Galileant invariant, which can be enforced paying attention to the particular nature of the D3Q19 lattice. A detailed derivation of the forcing term expressed in the velocity space, i.e., $F_i$, is provided in Appendix~\ref{app:AppForcingQ19}. Deriving it in the velocity space is of paramount importance because it allows its extension to any kind of moment space in a straightforward manner, as already demonstrated for both D2Q9 and D3Q27 lattices in our previous work~\cite{DEROSIS_PoF_31_2019}. Interestingly, accounting for external forces does not modify the rest of the procedure, which will henceforth be referred to as the D3Q19-CM-LBM. The interested reader may also refer to Appendix~\ref{app:AppRMQ19} for the raw moment (RM) formulation of the present algorithm. Moreover, the script D3Q19CentralMoments.m in the supplementary material allows the reader to perform all the symbolic manipulations to derive the proposed methodology.

\subsection{Some computational details} 
It is worth to highlight the benefits, in terms of computational cost and memory consumption, coming from the adoption of the present D3Q19-CM-LBM rather than the more standard D3Q27-CM-LBM~\cite{DEROSIS_PoF_31_2019}. Let us denote as $\Delta$ the number of lattice sites characterizing a certain LB simulation. The reduced memory requirements of the former model clearly stem when populations are considered. Indeed, one can save $(19/27) \times 100 \approx 30 \%$ when the simplest lattice model is considered (see Table~\ref{Table1}).
\begin{table}[tbp!]
\centering
\begin{tabular}{c | c }
\hline
\hline
Discretization & $f_i$ \\
\hline
Q19        & $19 \times \Delta$ \\
Q27        & $27 \times \Delta$ \\
Saving   & $\sim 30 \%$  \\
\hline
\hline
\end{tabular}
\caption{Memory usage involved by the D3Q19-CM-LBM and D3Q27-CM-LBM within a generic LB run. The most complete discretization involves an additional cost of $\sim 30 \%$.}
\label{Table1}
\end{table}
\\
\indent Now, let us consider the number of involved floating point operations. Firstly, one can immediately observe that the computation of macroscopic variables~(\ref{macro}) and pre-collision CMs~(\ref{pre_collision_central_moments}) needs to span a different number of directions (19 vs. 27). Hence, the simplest discretization allows us to reduce of approximately $30 \%$ the computational cost involved in the computation of $\rho$, $\bm{u}$ and $k_{5 \ldots 9}$. Moreover, the computation of post-collision moments and populations is drastically lighter when the 19-velocities discretization is adopted. In fact, the D3Q19-CM-LBM needs to evaluate the expressions in Eqs.~(\ref{rawQ19},\ref{pdfQ19}). One can immediately observe that for the D3Q27-CM-LBM [see Eqs.~(\ref{pdfQ27}) and the script D3Q27CentralMoments.m attached to the Supplemental Material], a larger number of floating point operations is required, hence, drastically increasing the computational cost of the CM-LBM as compared to the present D3Q19 formulation.

In the next section, we compare the numerical properties of the present D3Q19-CM-LBM against its D3Q27 counterpart~\cite{DEROSIS_PoF_31_2019} for the simulation of multiphysics flows.
The interested reader can refer to Appendix~\ref{app:Q27} for further details regarding the D3Q27-CM-LBMs.

\section{Numerical tests}
\label{SEC:III}
We compare the numerical properties of the D3Q19-CM-LBM with those of its D3Q27 counterpart through eight well-defined consolidated benchmark tests. The first five problems focus on the simulation of single phase flows in absence of external forces:
\begin{itemize}
\item Taylor-Green vortex,
\item double shear layer,
\item lid-driven cavity,
\item dipole-wall collision,
\item three-dimensional Taylor-Green vortex.
\end{itemize}
The sixth one, i.e. Hartmann flow, introduce the Lorentz force in the resulting magnetohydrodynamic system. The interested reader can refer to the work by Dellar~\cite{dellar2002lattice} for further details regarding the computation of the magnetic field.\\
This section ends with two cases dealing with multiphase flows, i.e.,
\begin{itemize}
\item  a static bubble of a certain fluid immersed in another one is considered by means of the well-known Shan-Chen pseudopotential force;
\item Rayleigh-Taylor instability mechanism is simulated hereafter by adopting the color-gradient method.
\end{itemize}
If not otherwise stated, populations will be initialized by assuming they are at equilibrium (the latter being computed thanks to initial macroscopic fields) and boundary conditions are imposed by the regularized technique~\cite{latt2008straight}.

\subsection{Taylor-Green vortex}

\begin{figure}[tbp!]
\centering
\subfigure{\includegraphics[trim={0 1cm 0 0},clip, width=0.45\textwidth]{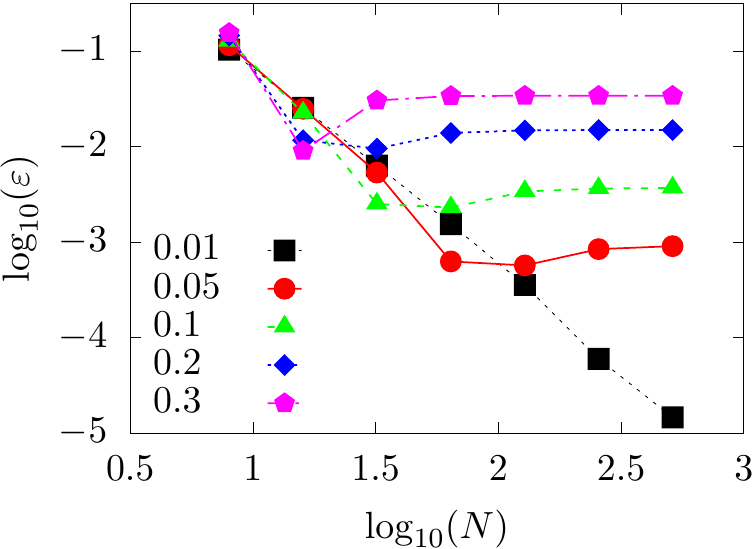}\label{Figure1a}}
\subfigure{\includegraphics[width=0.45\textwidth]{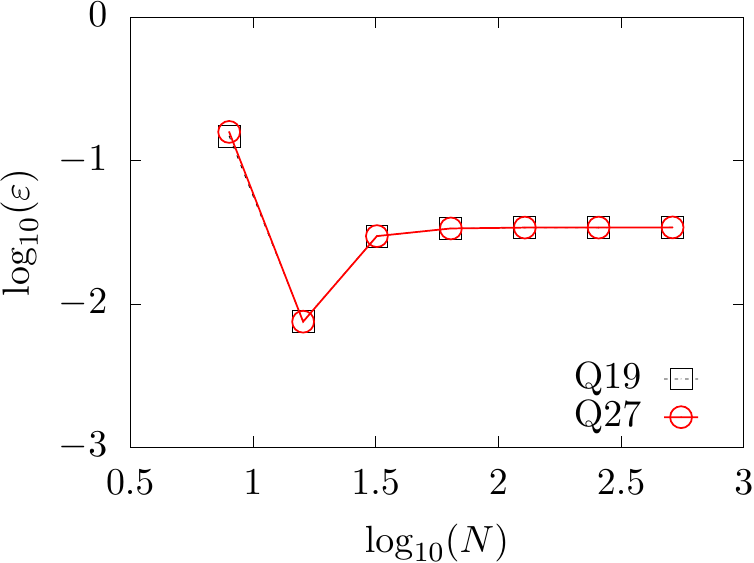}\label{Figure1b}}
\caption{Taylor-Green vortex: (top) convergence analysis carried out by setting $u_0=0.01$ (black filled squares), $0.05$ (red filled circles), $0.1$ (green filled triangles), $0.2$ (blue filled diamonds), $0.3$ (magenta filled pentagons), and (bottom) comparison between the results obtained by the D3Q19-CM-LBM (black squares) and D3Q27-CM-LBM (red circles) at $u_0=0.3$.}
\label{Figure1}
\end{figure}

We test the convergence properties of the adopted approach against the popular Taylor-Green vortex benchmark problem \cite{taylor1937mechanism}. Let us consider a square periodic domain of length $2 \pi$ with the following initial conditions:
\begin{eqnarray}
\rho \left(\bm{x},0 \right) &=& 3\rho_0 \left[1-\frac{3u_0^2}{4} \left(  \cos \left(2\xi x \right) +  \cos \left(2\xi y \right)  \right)      \right],  \nonumber \\
\bm{u} \left(\bm{x},0 \right)  &=& u_0 \left[\cos \left(\xi x \right) \sin \left(\xi y \right) , \,  \sin \left(\xi x \right) \cos \left(\xi y \right), \, 0 \right], 
\end{eqnarray}
with $\xi=2\pi/N$. The domain is idealized by $N \times N$ grid points in the $x-y$ plane, whereas only 1 point is adopted in the $z$ direction. The time evolution of the fluid velocity computed by our algorithm is compared to the analytical prediction
\begin{equation}
\bm{u} \left(\bm{x},t \right)  = \bm{u} \left(\bm{x},0 \right)\mathrm{exp}^{-t/T},
\end{equation}
where the characteristic time is $T = \left(2\xi^2 \nu \right)^{-1}$. Specifically, the numerical and analytical solutions are collected in the vectors $\boldsymbol \sigma^n$ and $\boldsymbol \sigma^a$, respectively, at $t=T$. Then, the relative error between the two is evaluated as
\begin{equation}\label{error}
\varepsilon = \frac{\| \boldsymbol \sigma^a- \boldsymbol \sigma^n  \|}{\| \boldsymbol \sigma^a \|},
\end{equation}
$\| \bullet \|$ denoting the L2-norm. A convergence analysis is carried out by varying the number of lattice points, $N$, discretizing each side of the domain, i.e. $N = 8,\, 16,\, 32,\, 64,\, 128, \, 256, \, 512$. Moreover, we investigate the influence of the Mach number by using four values of $u_0$, i.e. $u_0= 0.01,\,0.05,\,0.1,\,0.2,\,0.3$. We also set $\displaystyle \mathrm{Re} = \frac{u_0 N}{\nu}=1000$. Results are reported in Figure\ref{Figure1a}. For the lowest value of $u_0$, an optimal convergence value equal to 2 is found. As $u_0$ increases, the accuracy and convergence properties of the method deteriorate as well. This behavior should be addressed to the impossibility to add the diagonal terms $u_x^3$, $u_y^3$ and $u_z^3$ to the components of the third moments of the equilibrium populations~\cite{dellar2014lattice}.
These findings are consistent with those obtained by the D3Q27 lattice discretization. Figure~\ref{Figure1b} shows the results of the convergence analysis at $u_0=0.3$ by adopting the D3Q19- and D3Q27-CM-LBMs. Except for the coarsest resolution, curves are well overlapped, meaning that the accuracy loss observed for higher values of $u_0$ is not related the reduction of discrete velocities from 27 to 19. In fact, for both cases, the aliasing defect $c_{i\zeta}^3=c_{i\zeta}$ ($\zeta=x$, $y$ or $z$) is at the origin of the velocity-dependent error terms, that are related to compression/dilation phenomena (trace of the viscous stress tensor), and which can only be dealt with using correction terms~\cite{DELLAR_JCP_259_2014,HOSSEINI_RSTA_378_2020,RENARD_ARXIV_2020_08477}.

\subsection{Double shear layer}
\label{DoubleShearLayer}

\begin{figure}[tbp!]
\centering
\includegraphics[width=0.45\textwidth]{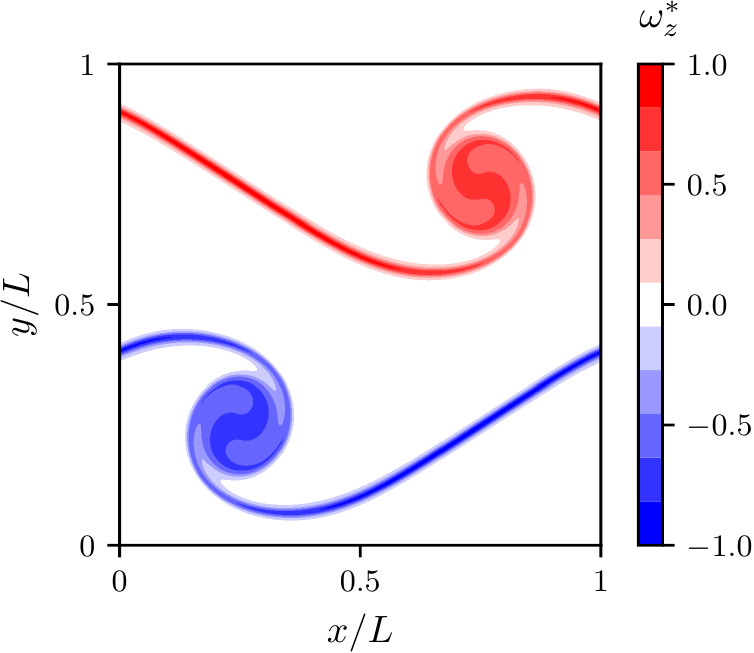}
\caption{Double shear layer: normalized vorticity field $\omega_z^*=\omega_z/\omega_z^{\mathrm{max}}$ at $t/t_0=1$ showing the roll-up of the shear layers, and the generation of two counter-rotating vortices.}
\label{Figure2}
\end{figure}

\begin{figure}[tbp!]
\centering
\subfigure{\includegraphics[trim={0 1.1cm 0 0},clip, width=0.49\textwidth]{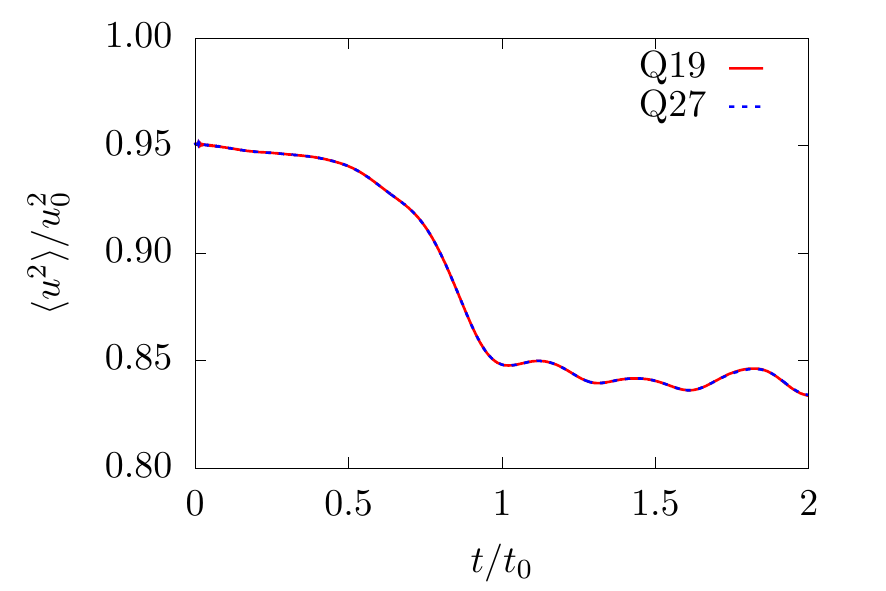}\label{Figure3a}}
\subfigure{\includegraphics[width=0.49\textwidth]{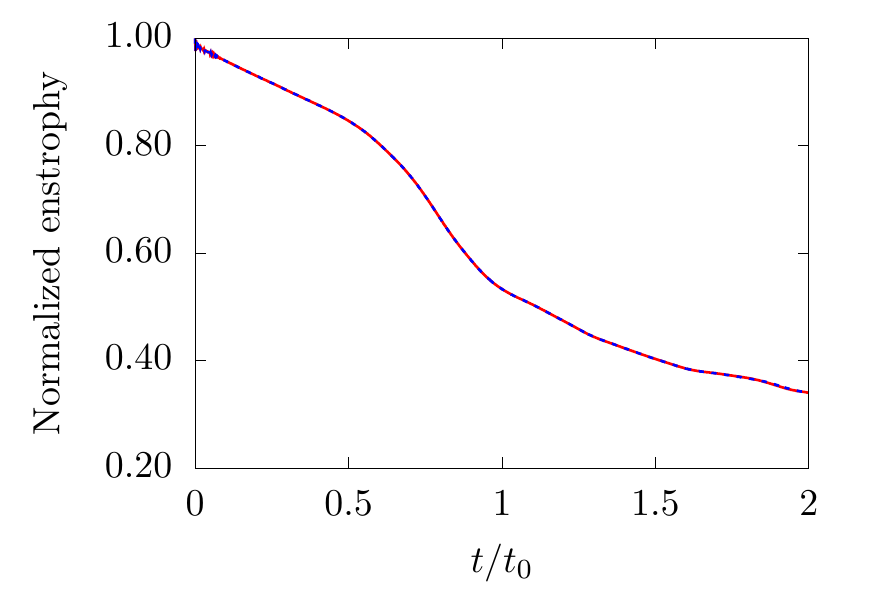}\label{Figure3b}}
\caption{Double shear layer: evolution of (top) the normalized kinetic energy and (bottom) normalized enstrophy by D3Q19-CM-LBM (red continuous line) and D3Q27-CM-LBM (blue dotted line). The two approaches generate results that are very well overlapped.}
\label{Figure3}
\end{figure}

\begin{figure*}[htbp!]
\centering
\subfigure{\includegraphics[trim= 0 1.1cm 0 0.2cm, clip, width=0.45\textwidth]{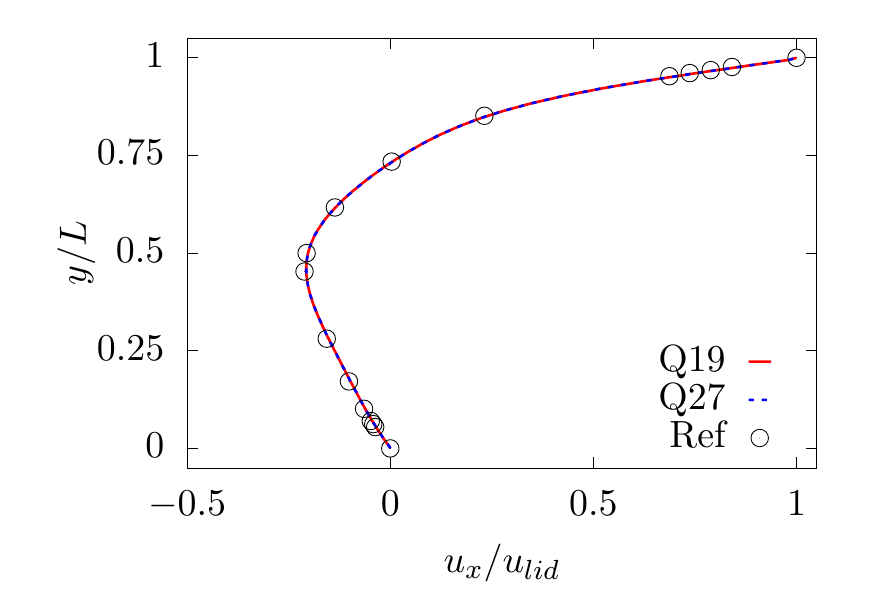}}
\subfigure{\includegraphics[trim= 0 1.1cm 0 0.2cm, clip, width=0.45\textwidth]{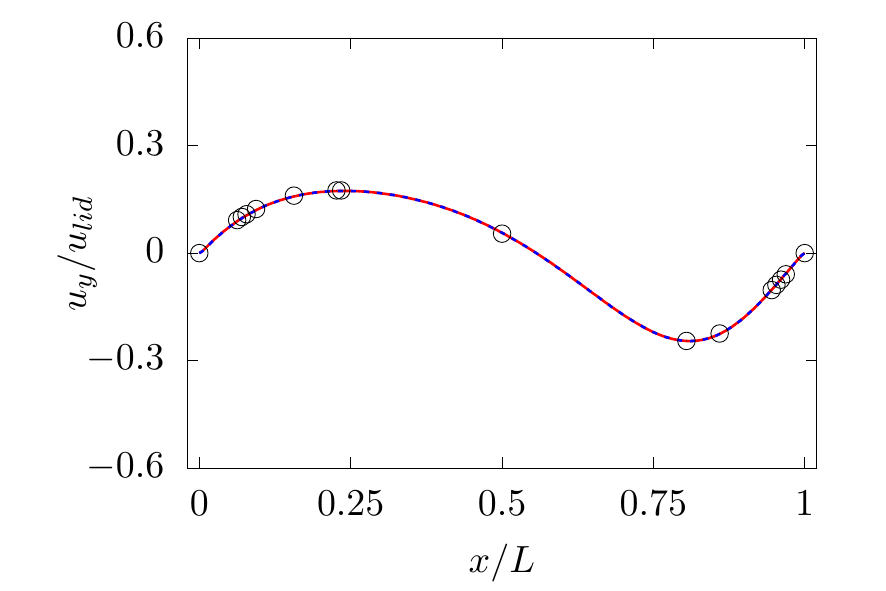}}
\subfigure{\includegraphics[trim= 0 1.1cm 0 0.2cm, clip, width=0.45\textwidth]{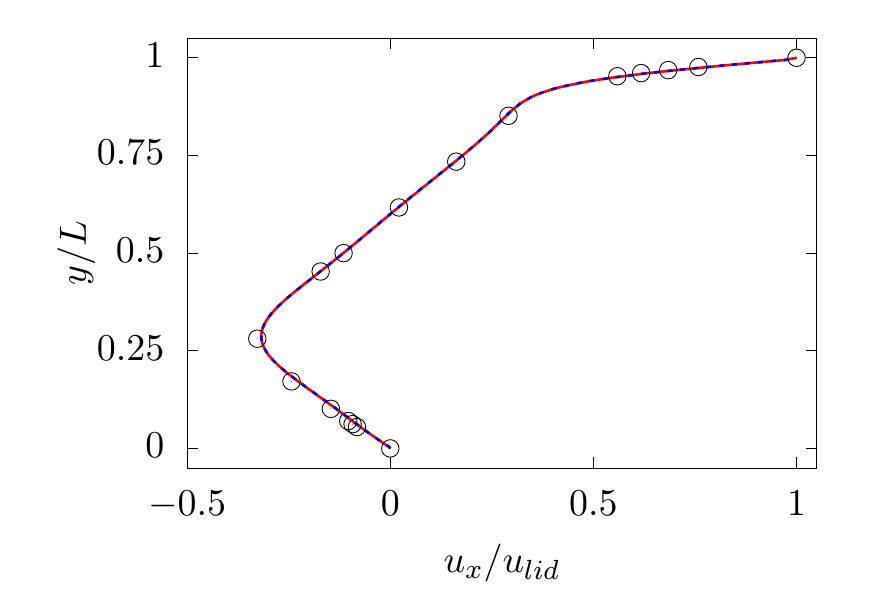}}
\subfigure{\includegraphics[trim= 0 1.1cm 0 0.2cm, clip, width=0.45\textwidth]{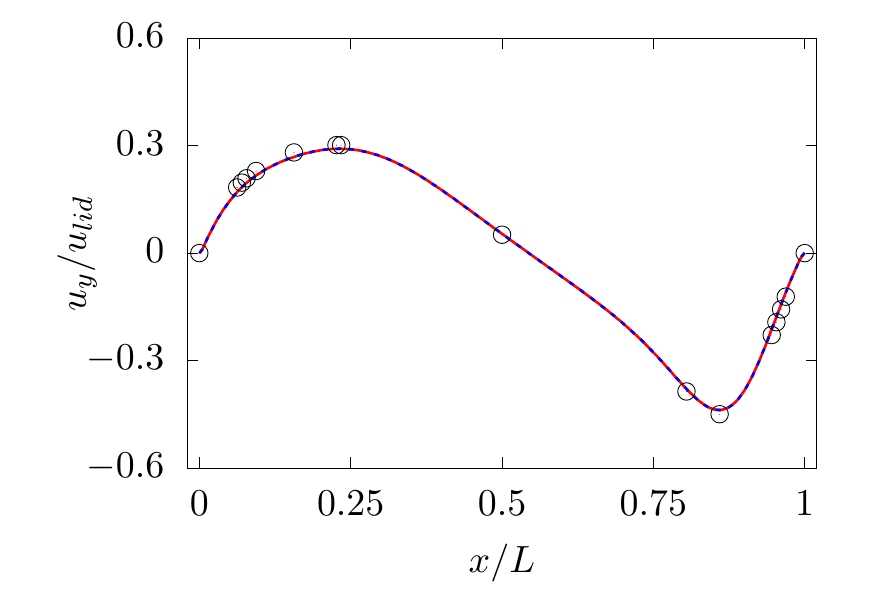}}
\subfigure{\includegraphics[trim= 0 0.1cm 0 0.2cm, clip, width=0.45\textwidth]{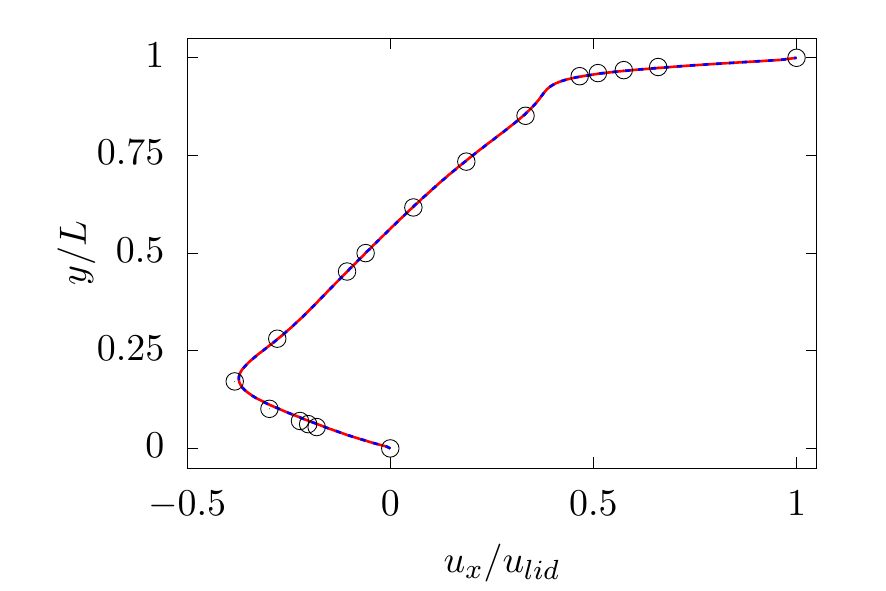}}
\subfigure{\includegraphics[trim= 0 0.1cm 0 0.2cm, clip, width=0.45\textwidth]{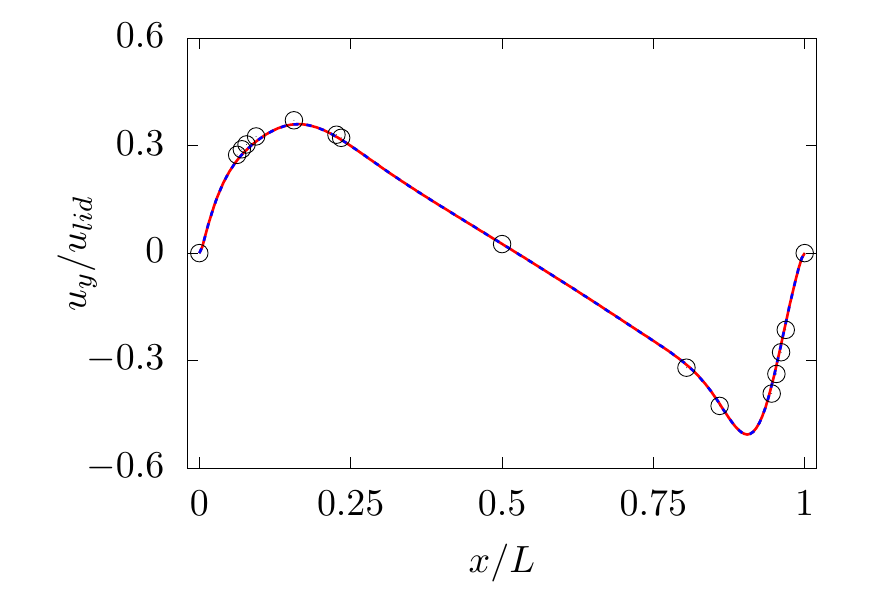}}
\caption{Lid-driven cavity: profiles of the horizontal component of the velocity along the vertical mid-section (left) and profiles of the vertical component of the velocity along the horizontal mid-section (right) at Re=100 (top), 400 (mid) and 1000 (bottom) by D3Q19-CM-LBM (red continuous line) and D3Q27-CM-LBM (blue dotted line). A very good agreements with reference values (Ref) in Ref.~\cite{ghia1982high} is found.}
\label{Figure4}
\end{figure*}

An excellent candidate to evaluate the stability of any numerical scheme is represented by the double shear layer test \cite{BROWN_JCP_122_1995,MINION_JCP_138_1997}. By considering a two-dimensional periodic domain with $(x,y) \in [0,L]^2$, initial conditions are given by two longitudinal shear layers and a superimposed transverse perturbation, i.e.,
\begin{equation}
u_x(\bm{x},t=0) = \left\{
			\begin{array}{ll}
			u_0\, \mathrm{tanh}\left[\kappa \left(\frac{y}{L}- \frac{1}{4} \right) \right], \qquad \frac{y}{L} \leq \frac{1}{2},\\
			\\
			u_0\, \mathrm{tanh}\left[\kappa \left(\frac{3}{4} -\frac{y}{L} \right) \right], \qquad \frac{y}{L} > \frac{1}{2},
			\end{array}
              \right.
\end{equation}
and
\begin{equation}
u_y(\bm{x},t=0) = u_0 \delta \sin \left[2 \pi \left( \frac{x}{L}+\frac{1}{4} \right) \right],
\end{equation}
where $\kappa=80$ and $\delta=0.05$. The Reynolds and Mach numbers are $\mathrm{Re} = u_0 L / \nu=3 \times 10^4$ and $\mathrm{Ma}=u_0/c_s=0.57$, respectively, with $L=256$. Only one point is considered in the direction $z$. \\
\indent Figure~\ref{Figure2} sketches the \textcolor{black}{(normalized)} vorticity field at $t/t_0=1$ (with $t_0=L/u_0$) and confirms the rise of a Kelvin-Helmholtz instability mechanism, where the flow physics manifests the roll-up of the shear layers and the generation of two counter-rotating vortices.\\
\indent In Figure~\ref{Figure3a}, the time history of the kinetic energy (normalized by its initial value) is reported by adopting the D3Q19-CM-LBM and D3Q27-CM-LBM. As it stems from the graph, results are in excellent agreement with a percentage relative discrepancy of $\sim 0.009 \%$. These findings are confirmed by the plot of the kinetic enstrophy (normalized by the initial value) in Figure~\ref{Figure3b}, where curves obtained by the two approaches are, again, overlapped.\\
\indent In Appendix~\ref{app:AppExtendedEquilibrium}, this test is further \textcolor{black}{used to demonstrate the stability improvement induced by the extended equilibrium~(\ref{eq:EqQ19RM}) as compared to its second-order counterpart}.

\begin{figure*}[tbp!]
\centering
\includegraphics[width=0.96\textwidth]{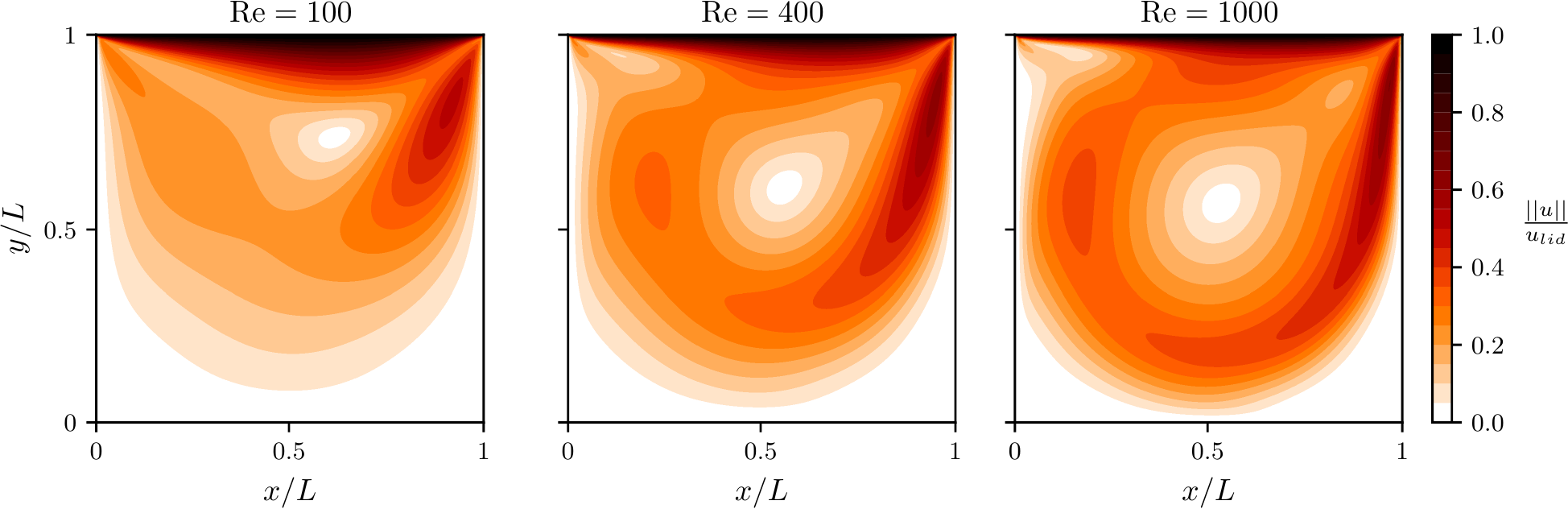}
\caption{Lid-driven cavity: map of normalized velocity at different Reynolds number, and for a grid mesh composed of $L=100$ points in each direction. }
\label{Figure5}
\end{figure*}

\subsection{Lid-driven cavity}

The lid-diven cavity~\cite{ghia1982high, botella1998benchmark} represents one of the most canonical problem to evaluate the accuracy of numerical schemes. Let us consider a square domain of length $L=201$. At the top section, a constant uniform rightward velocity $u_{lid}=0.01$ is imposed, while the no-slip condition is enforced at the remaining edges. The initial conditions are $\rho(\bm{x},t=0)=1$ and $\bm{u}(\bm{x},t=0)=0$. Figure~\ref{Figure4} sketches the velocity profiles in the two mid-sections for different values of the Reynolds number, i.e. $\displaystyle \mathrm{Re} = \frac{u_{lid} L}{\nu}=100,\, 400,\, 1000$. Findings obtained by the D3Q19-CM-LBM are very well-overlapped to those provided by the D3Q27-CM-LBM, that, in turn, exhibit a very good match with the reference ones in Ref.~\cite{ghia1982high}, thus highlighting the accuracy of the proposed approach. {\color{black}It is of note that Ghia et al.~\cite{ghia1982high} formulated the governing equations in vorticity-stream function variables and used a strongly coupled implicit multigrid to solve this problem for $\displaystyle \mathrm{Re} \leq 10000$.}
For the sake of completeness, the velocity field at the end of each simulation is reported in Figure~\ref{Figure5}. Again, the contour plot is in full agreement with those drawn in Ref.~\cite{ghia1982high}.

\subsection{Dipole-wall collision}

\begin{table*}[tbp!]
\centering
\begin{tabular}{ c | c | c | c | c | c | c | c | c | c  | c | c }
\hline
\hline
$\mathrm{Re}$ & $t$ & $E_{Q19}$ & $E_{Q27}$ & $E$~\cite{MOHAMMED201879} & $E_{FD}$~\cite{CLERCX2006245} & $E_{SM}$ \cite{CLERCX2006245} & $\Psi_{Q19}$ & $\Psi_{Q27}$ & $\Psi$ \cite{MOHAMMED201879} &  $\Psi_{FD} $ \cite{CLERCX2006245} &  $\Psi_{SM} $ \cite{CLERCX2006245}\\
\hline
\multirow{3}{*}{625} & 0.25 & 1.494 & 1.494 & 1.501 & 1.502 & 1.502   & 467.2 & 467.2 & 472.1  & 472.7  & 472.6\\
                                    & 0.5   & 1.010 & 1.010 & 1.013 & 1.013 & 1.013    & 374.0 & 374.0 & 382.6 & 380.6 & 380.6\\
                                    & 0.75 & 0.765 & 0.765 & 0.767 & 0.767 & 0.767 & 244.8 & 244.8 & 256.0 & 255.0 & 255.2\\
\hline
\multirow{3}{*}{1250} & 0.25 & 1.710 & 1.710 & 1.719 & 1.721 & 1.720 & 603.6 & 603.6 & 613.6  & 615.0 & 615.0\\
                                     & 0.5   & 1.308 & 1.308 & 1.312 & 1.313 & 1.313 & 601.7 & 601.6 & 612.8  & 611.3  & 611.9\\
                                     & 0.75 & 1.057 & 1.057 & 1.061 & 1.061 & 1.061 & 473.1 & 473.2 & 486.2 & 484.4 & 484.7\\
\hline
\multirow{3}{*}{2500} & 0.25 & 1.838 & 1.838 & 1.848 & 1.851 & 1.850 & 705.3 & 705.3 & 725.6  & 727.8 & 728.2\\
                                     & 0.5   & 1.534 & 1.534 & 1.540  & 1.541 & 1.541  & 898.1 & 898.0 & 917.6  & 916.6  & 920.5\\
                                     & 0.75 & 1.320 & 1.320 & 1.325  & 1.326 & 1.326 & 790.2 & 790.1 & 809.9 & 805.5 & 808.1\\
\hline	
\hline
\end{tabular}
\caption{Normal dipole-wall collision: energy $E$ and enstrophy $\Psi$ at salient time instants. Reproduced with permission from Comput. Fluids 176, (2018). Copyright 2018 Elsevier. Reproduced with permission from Comput. Fluids 35, (2006). Copyright 2006 Elsevier.}
\label{Table2}
\end{table*}

\begin{figure*}[tbp!]
\centering
\subfigure{\includegraphics[width=0.45\textwidth]{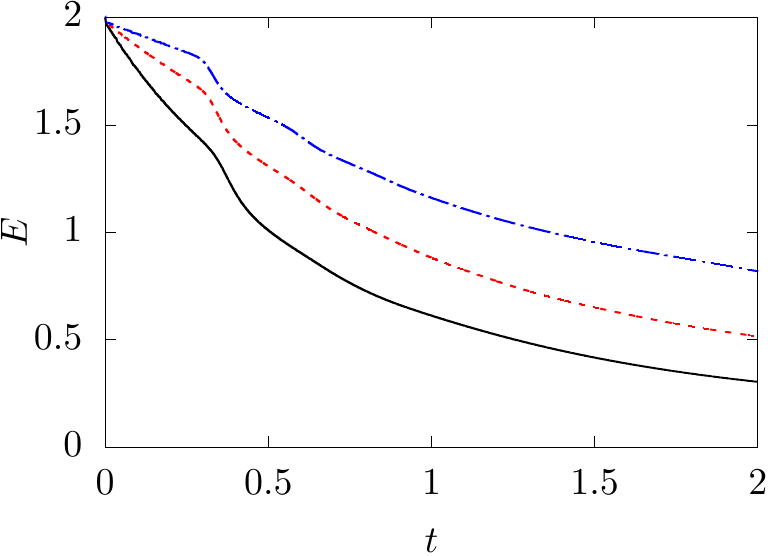}}\hspace*{1cm}
\subfigure{\includegraphics[width=0.45\textwidth]{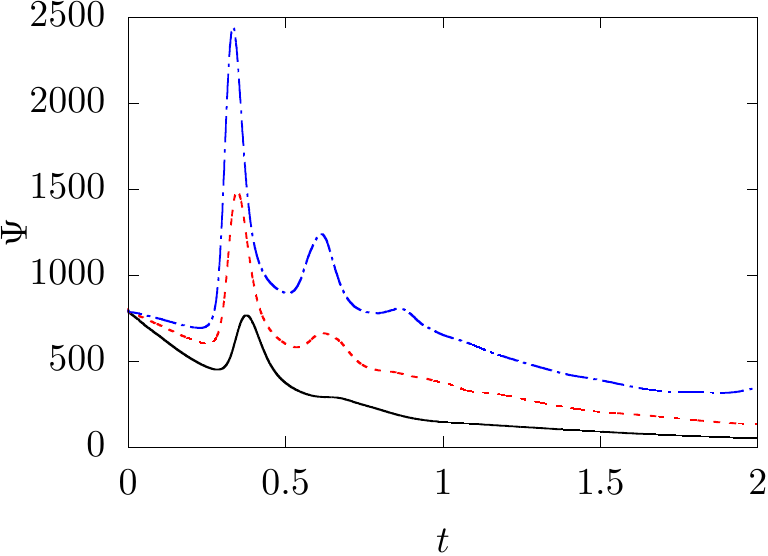}}
\caption{Normal dipole-wall collision: time evolution of the energy $E$ and enstrophy $\Psi$ at different Reynolds number: 625 (black continuous line), 1250 (red dashed line) and 2500 (blue dash-dotted line).}
\label{Figure6}
\end{figure*}

\begin{table}[tbp!]
\centering
\begin{tabular}{ c | c | c | c | c | c }
\hline
\hline
$\mathrm{Re}$ & $t$ & $E_{Q19}$ & $E_{Q27}$ & $E$ \cite{MOHAMMED201879}  & $\varepsilon\, [\%]$\\
\hline
\multirow{3}{*}{625} & 0.3 & 1.416 & 1.416 & 1.423 & 0.4\\
 & 0.5 & 1.049 & 1.049 & 1.049 & 0 \\
 & 2.0 & 0.391 & 0.391 & 0.386 & 1.2 \\
 \hline
 \multirow{3}{*}{1250} & 0.3 & 1.651 & 1.651 & 1.659 & 0.4 \\
 & 0.5 & 1.349 & 1.349  & 1.353 & 0.2\\
 & 2.0 & 0.680 & 0.680 & 0.675 & 0.7\\
 \hline
 \multirow{3}{*}{2500} & 0.3 & 1.793 & 1.793 & 1.790 & 0.1\\
 & 0.5 & 1.574 & 1.574 & 1.579 & 0.3\\
 & 2.0 & 1.043 &1.043 & 1.053 & 0.9\\
\hline
\hline
\end{tabular}
\caption{Inclined dipole-wall collision: kinetic energy computed by the proposed approach at different time instants and Reynolds number against recent findings in Ref.~\cite{MOHAMMED201879}. Reproduced with permission from Comput. Fluids 176, (2018). Copyright 2018 Elsevier.}
\label{Table3}
\end{table}

\begin{table*}[tbp!]
\centering
\begin{tabular}{ c | c | c | c | c | c | c | c | c }
\hline
\hline
$\mathrm{Re}$ & $t_1$(Q19) & $t_1$(Q27) &  $t_1$(\cite{MOHAMMED201879}) &  $\varepsilon\, [\%]$ & $t_2$(Q19) & $t_2$(Q27) &  $t_2$(\cite{MOHAMMED201879}) &  $\varepsilon\, [\%]$ \\
\hline
625 & 0.364 & 0.364 & 0.364 & 0 & 0.638 & 0.638 & 0.647 & 1.4\\
1250 & 0.343  & 0.343 & 0.333 & 3.0 & 0.583 & 0.583 & 0.583 & 0 \\
2500 & 0.328  & 0.328 & 0.325 & 0.9 & 0.570 & 0.570 & 0.570 & 0\\
\hline
\hline
\end{tabular}
\caption{Inclined dipole-wall collision: time instants when the first and second maximum of the enstrophy manifest. Reproduced with permission from Comput. Fluids 176, (2018). Copyright 2018 Elsevier.}
\label{Table4}
\end{table*}

We further evaluate the numerical performance of the D3Q19-CM-LBM by examining the flow physics generated by a dipole-wall collision \cite{CLERCX2006245, MOHAMMED201879}. Let us consider a square domain $(x,y) \in [-1:1]^2$, enclosed by no-slip walls at each side. The velocity is initialized as
\begin{widetext}
\begin{align}
u_x(\bm{x},t=0) &= -\frac{1}{2} | w_e | \left(y-y_1 \right) \mathrm{exp}\left[-\left( r_1/r_0\right)^2   \right] + \frac{1}{2} | w_e | \left(y-y_2 \right) \mathrm{exp}\left[-\left( r_2/r_0\right)^2   \right],\nonumber \\
u_y(\bm{x},t=0) &= \phantom{-}\frac{1}{2} | w_e | \left(x-x_1 \right) \mathrm{exp}\left[-\left( r_1/r_0\right)^2   \right] - \frac{1}{2} | w_e | \left(x-x_2 \right) \mathrm{exp}\left[-\left( r_2/r_0\right)^2   \right],
\end{align}
\end{widetext}
where the positions of the two monopoles are $(x_1, \, y_1) = (0, \, 0.1)$ and $(x_2, \, y_2) = (0, \, -0.1)$. Their radius is $r_0=0.1$, $r_{\alpha} = \sqrt{\left(x-x_{\alpha} \right)^2+\left(y-y_{\alpha} \right)^2}$ (with $\alpha=1,2$) and the strength of the monopoles is $w_e=299.56$. Under this setup, the initial values of the kinetic energy and enstrophy are
\begin{eqnarray}
E(t=0) &=& \frac{1}{2} \int_{-1}^{1} \int_{-1}^{1} | \bm{u}^2 |\left( \bm{x},t=0 \right) \, \mathrm{d}x  \, \mathrm{d}y = 2, \nonumber \\
\Psi(t=0) &=& \frac{1}{2} \int_{-1}^{1} \int_{-1}^{1} | \psi^2 |\left( \bm{x},t=0 \right) \, \mathrm{d}x  \, \mathrm{d}y = 800, \label{dipole}
\end{eqnarray}
respectively, with $\psi = \partial_x u_y - \partial_y u_x$. The characteristic Reynolds number is $\mathrm{Re} = (UD)/\nu$, where $U=1$ is the root-mean-square of the velocity field in Eqs.~(\ref{dipole}) and $D=1$ is the half width of the domain. Before performing any LB run, a proper number of lattice sites to discretize the domain should be chosen. Following the grid independence analysis in \cite{MOHAMMED201879}, we adopt $D_{lb}=512$, $768$ and $1024$ to simulate scenarios at $\mathrm{Re}=625$, 1250 and 2500, respectively. The latter grid meshes ensure the proper resolution of all features of the flow, at least, in the normal collision configuration. Notice that $D_{lb}$ is the number of points idealizing $D$. Our numerical simulations are carried out at a Mach number $\mathrm{Ma} \sim 0.06$.

In Table~\ref{Table2}, the values of the energy $E$ and enstrophy $\Psi$ at salient time instants are reported. Findings from the D3Q19-CM-LBM run are compared to the reference solution in Ref.~\cite{CLERCX2006245} and to a recent LB effort \cite{MOHAMMED201879}.

One can immediately observe that the D3Q19- and the D3Q27-CM-LBMs produce identical results. In turn, they show a slight mismatch (up to $3\%$) with respect to the LB study by Mohammed \textit{et al.}~\cite{MOHAMMED201879}. It should be noted that findings in Ref.~\cite{MOHAMMED201879} are closer to the reference ones by Clercx \& Bruneau~\cite{CLERCX2006245}. We address this behavior to the adoption in Ref.~\cite{MOHAMMED201879} of (i) a more accurate boundary condition and (ii) a lower Mach number. The time evolutions of the energy and enstrophy are reported in Figure~\ref{Figure6}.


Furthermore, we investigate a configuration where an inclined collision is present. Specifically, we rotate the dipole by 30 degrees counter-clockwise by setting $(x_1, \, y_1) = (0.0839, \, 0.0866)$ and $(x_2, \, y_2) = (0.1839, \, -0.0866)$. In Table~\ref{Table3} we compare the kinetic energy computed by the proposed approach at different time instants and Reynolds number against recent findings in Ref.~\cite{MOHAMMED201879}. Again, a very good agreement is found with a slight mismatch up to $1.2\%$. The accuracy of the method is further highlighted in Table~\ref{Table4}, where the time instants corresponding to the rise of the the first and second maxima of the enstrophy agree very well with those in Ref.~\cite{MOHAMMED201879}.

\begin{figure*}[tbp!]
\centering
\subfigure[$\:\text{Normal collision}$.]{\includegraphics[width=0.99\textwidth]{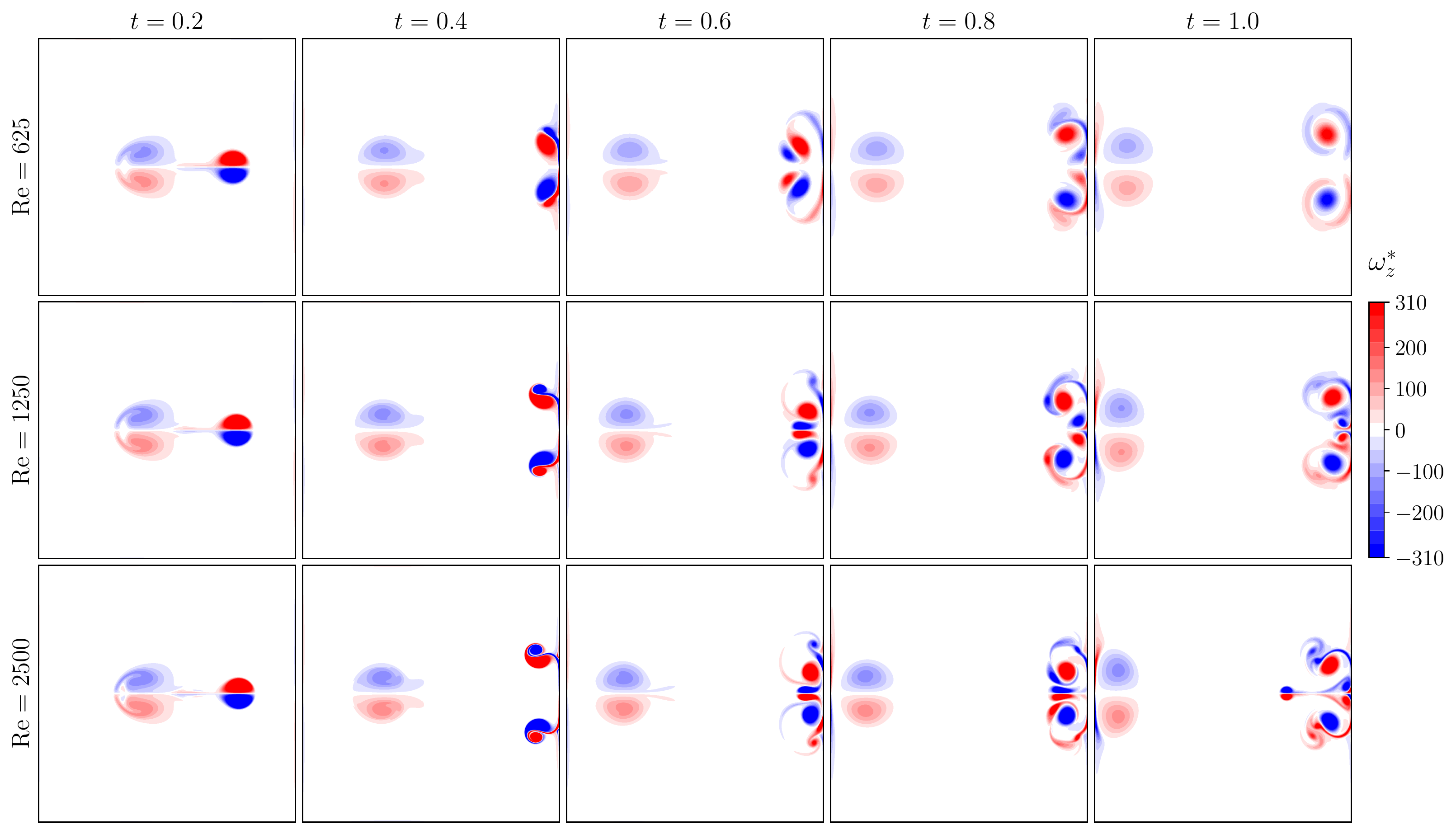}}
\subfigure[$\:\text{Inclined collision}$.]{\includegraphics[width=0.99\textwidth]{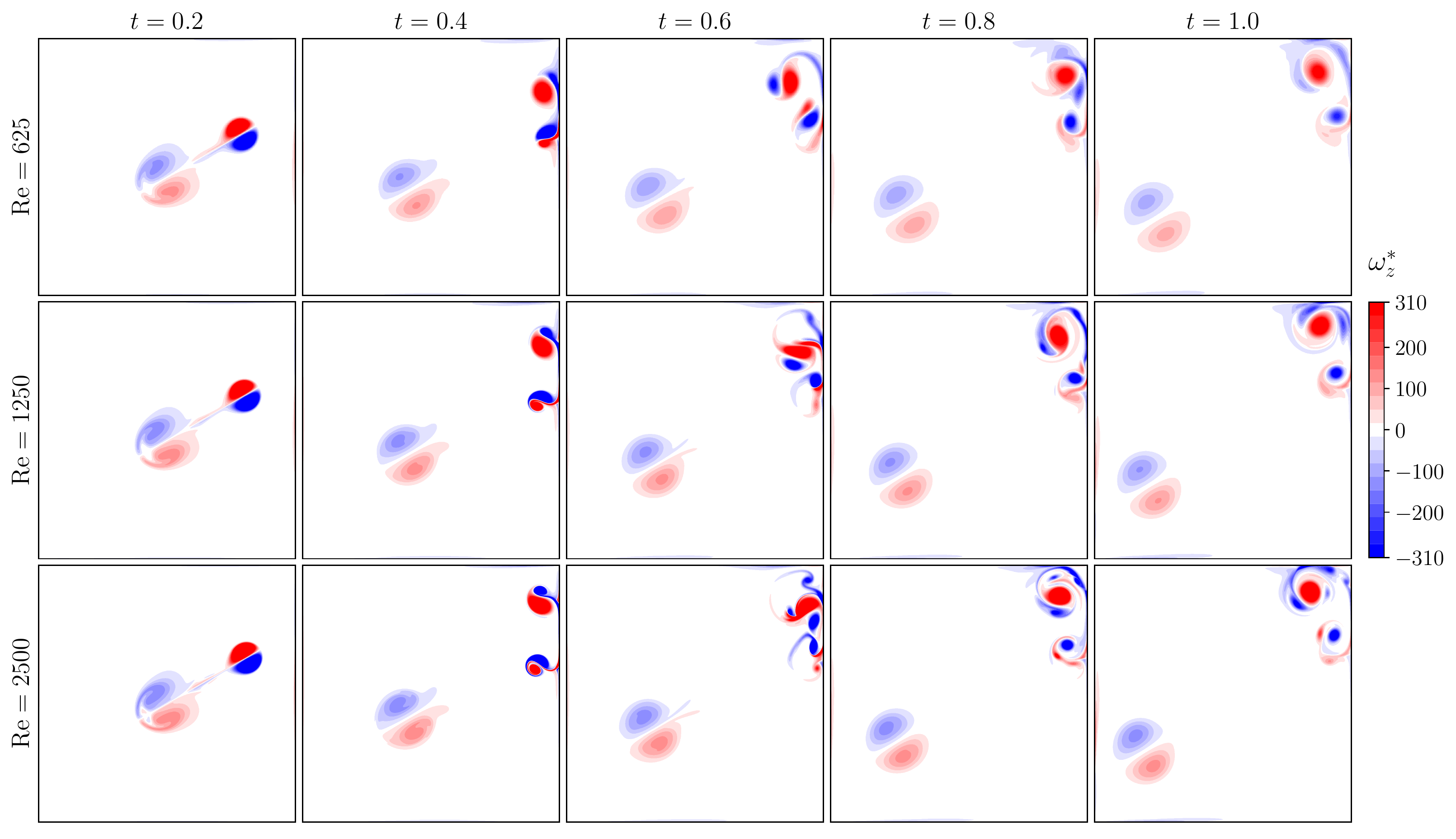}}
\caption{Dipole-wall collision: vorticity map at salient time instants and different Reynolds numbers. Normal (a) and inclined (b) collisions are sketched.}
\label{Figure7}
\end{figure*}

In Figure~\ref{Figure7}, the vorticity magnitude is sketched at salient time instants for the afore-mentioned configuration. Present findings corroborate those in Ref.~\cite{MOHAMMED201879}. In particular, both studies show that the vortex hits the left wall at $t=1$ if the normal collision is considered, with progressively smaller-scale structures arising as $\mathrm{Re}$ increases.

\subsection{Three-dimensional Taylor-Green vortex}
\label{3DTG}

\begin{figure*}[tbp!]
\centering
\subfigure[$\: \mathrm{Re} =1600$.]{\includegraphics[width=0.49\textwidth]{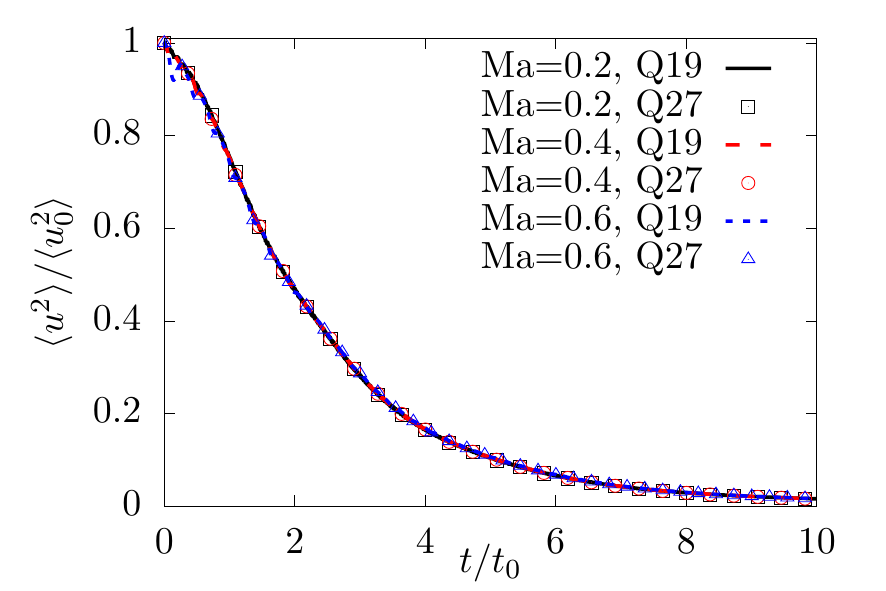}}
\subfigure[$\: \mathrm{Re} =30000$.]{\includegraphics[width=0.49\textwidth]{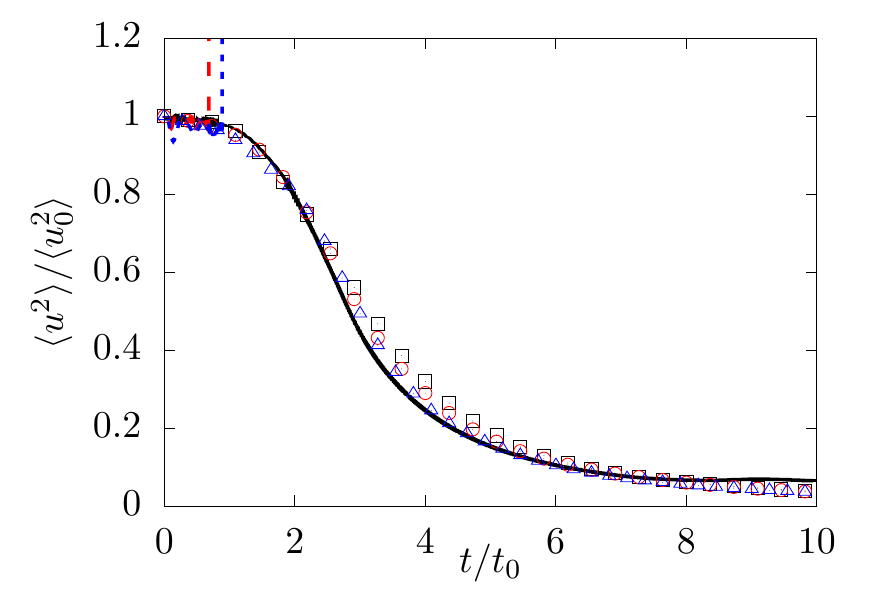}}
\caption{Taylor-Green vortex: time evolution of the kinetic energy normalized by its initial value. Lines and symbols denote to the adoption of the D3Q19-CM-LBM and D3Q27-CM-LBM, respectively. Findings correspond to $\mathrm{Ma} = 0.2$ (black solid line and squares), 0.4 (red dashed line and circles), 0.6 (blue dotted line and circles).}
\label{Figure8}
\end{figure*}

\begin{figure*}[tbp!]
\centering
\subfigure{\includegraphics[width=0.95\textwidth]{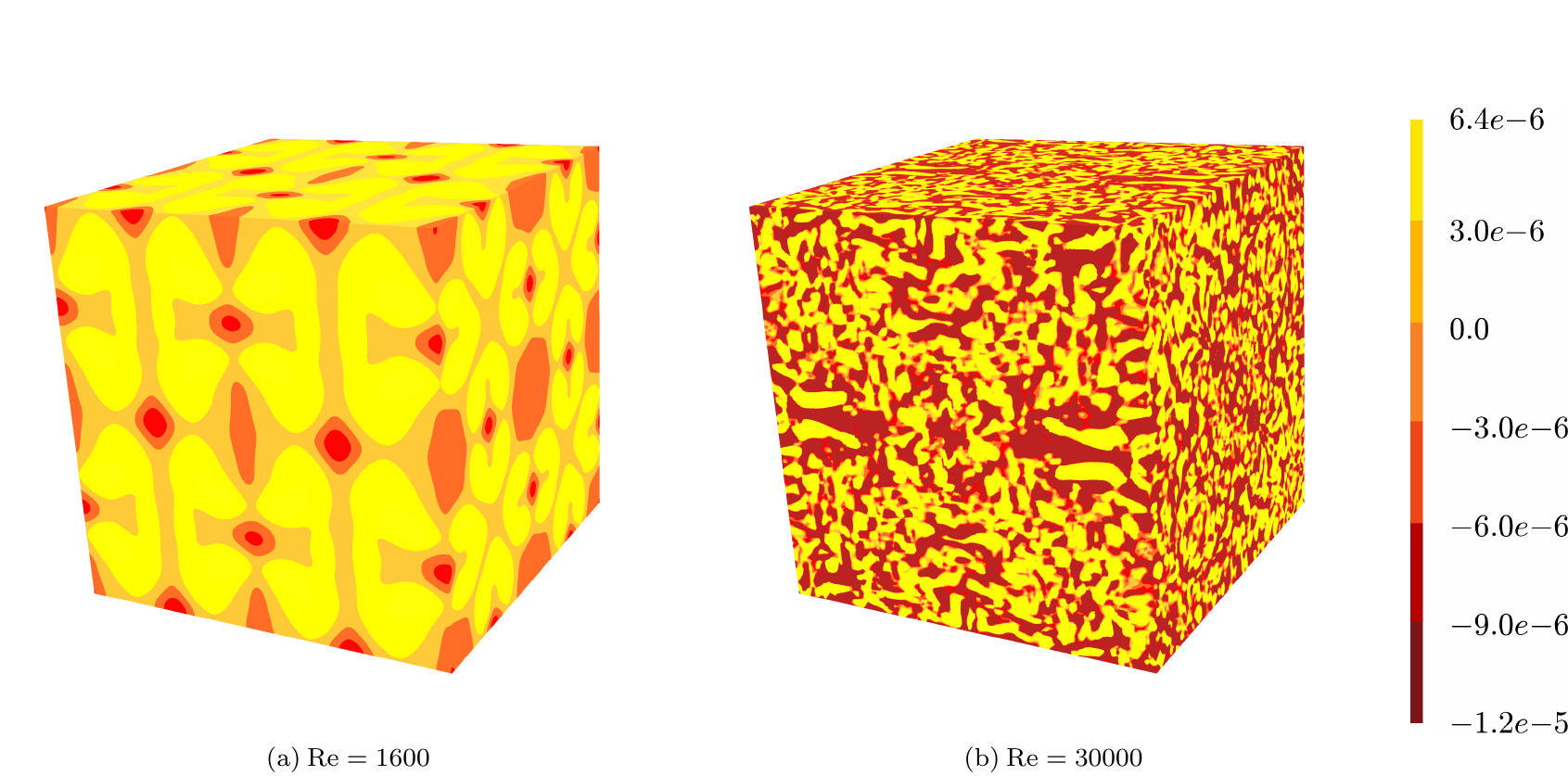}}
\caption{Taylor-Green vortex: map of the Q-criterion at $t=5$. Findings correspond to $\mathrm{Ma} = 0.2$.}
\label{Figure9}
\end{figure*}

We investigate the numerical performance of the proposed methodology against a three-dimensional Taylor-Green vortex \citep{aubard2013comparison,kajzer2014large}. Let us consider a cubic periodic domain with edge length $D$. The flow develops due to the following initial conditions:
\begin{eqnarray}
u_x(\bm{x},t=0) &=& u_0 \cos x \sin y \sin z, \nonumber \\
u_y(\bm{x},t=0) &=& -\frac{u_0}{2} \sin x \cos y \sin z, \nonumber \\
u_z(\bm{x},t=0) &=& -\frac{u_0}{2} \sin x \sin y \cos z.
\end{eqnarray}
By setting $D=128$, we run several simulations by varying the Reynolds number $\mathrm{Re} = u_0 D/ \nu=1600, \, 30000$ and Mach number $\mathrm{Ma} = u_0 /c_s = 0.2, \, 0.4, \, 0.6$. In Figure~\ref{Figure8}, results from all the above mentioned cases are reported in terms of the evolution of the kinetic energy normalized by its initial value. For the lowest value of $\mathrm{Re}$, findings are substantially insensitive to $\mathrm{Ma}$. Moreover, results obtained by the adoption of the D3Q19-CM-LBM overlap very well those provided by its D3Q27 counterpart, with a relative discrepancy of $\sim 0.8\%$. However, the behavior becomes more interesting when a higher value of $\mathrm{Re}$ is considered. Indeed, the adoption of the D3Q19-CM-LBM leads to a stable simulation only 
for $\mathrm{Ma}=0.2$, where diffusive phenomena seem to be underestimated due to remaining velocity-dependent errors in the viscous stress tensor. The latter issue eventually leads to stability issues, at $t/t_0 \sim 1$ with $t_0=D/u_0$, for higher Mach numbers. On the other hand, the D3Q27-CM-LBM does not undergo any instability. Notably, the relative difference between the solutions provided by the two algorithms is now more prominent and equal to $\sim 7.9 \%$.

Finally, the second invariant of the velocity gradient tensor, also know as Q-criterion, is depicted in Figure~\ref{Figure9} at $t=5$, where the turbulent behavior of the flow can be appreciated especially at $\mathrm{Re}=30000$.
An animation of the vorticity field is also available at \protect\url{https://www.youtube.com/watch?v=QfQ_CpN1CV4}, together with one of the Q-criterion \protect\url{https://www.youtube.com/watch?v=OKTh4YWjZ6g}. As a conclusion, this testcase shows that one condition to move from the D3Q27 formulation to its D3Q19 counterpart would be to ensure that $\mathrm{Ma}\leq 0.2$ for under-resolved conditions, in order to keep good stability and accuracy properties.

\subsection{Hartmann flow}
In order to test the capability of the present model with forcing, we investigate the so-called Hartmann flow, that is the analogous of the Poiseuille flow for an electrically conductive fluid of magnetic resistivity $\eta=\nu$. Here, the forcing scheme (\ref{eq:posColl0}) is based on the new formulation (\ref{eq:FRMQ19}). The latter is used to account for the Lorentz force $\bm{F}= \bm{j} \times \bm{b}$, $\bm{j}$, where $\bm{j}$ is the electric current that is computed directly from the populations~\cite{pattison2008progress} and $\bm{b}$ is the magnetic field. Let us assume a rectangular domain height $L$. Initial conditions consist of $\bm{b}(\bm{x}, t=0) = [0, b_{y0}, 0]$. The channel is periodic in the horizontal direction, while a constant uniform vertical magnetic field ($b_{y0}$) is enforced at the bottom and top walls, where the no-slip condition is enforced too. The Hartmann flow admits analytical solution in the form:
\begin{equation} \label{imposed_hartman}
u_x(\bm{x},t) =\frac{4 \nu u_0}{L b_{y0} \tanh \left(\mathrm{Ha} \right)} \left[ 1-\frac{\cosh \left(\mathrm{Ha} y^{\prime}/L \right)}{\cosh \left(\mathrm{Ha} \right)}    \right],
\end{equation}
where $y^{\prime}=2y-L$ and the Hartmann number is defined as $\mathrm{Ha} = b_{y0} L/\sqrt{4\rho_0 \nu \eta}$~\cite{dellar2002lattice}. 

\begin{figure}[tbp!]
\centering
\includegraphics[scale=1]{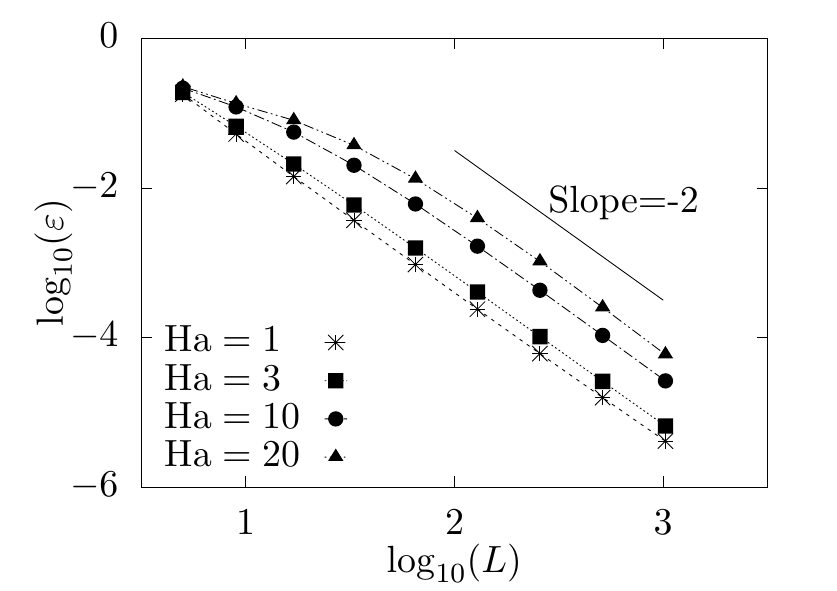}
\caption{Hartmann flow simulation: convergence analysis for $\mathrm{Ha}=1$ (stars, dashed line), 3 (squares, dotted line), 10 (circles, dot-dashed line) and 20 (triangles, dot-dot-dashed line). The continuous line denotes a convergence rate equal to 2.}
\label{Figure10}
\end{figure}

The convergence properties of the proposed approach are sketched in Figure~\ref{Figure10}, where the relative discrepancy between the computed numerical solution and the analytical one is plotted against the number of points representing the vertical dimension of the channel ($L \in [9:1025]$). Present results are in full agreement with those in De Rosis \textit{et al.}~\cite{DEROSIS_PoF_31_2019}, where an optimal convergence rate equal to 2 is found. Moreover, and again in agreement with Ref.~\cite{DEROSIS_PoF_31_2019}, a poor convergence is experienced for low values of $L$ and high values of $\mathrm{Ha}$, where the presence of thin Hartmann layers requires a larger number of grid points to be successfully and accurately reconstructed. 

The accuracy of the proposed approach is compared to the solution provided by the D3Q27-CM-LBM. Specifically, we re-run the simulation at $\mathrm{Ha=20}$ by using the finer lattice discretization and the results are summarized in Table~\ref{TableHartmann}. One can immediately appreciate that the two schemes exhibit very similar accuracy.
\begin{table*}[!htbp]
\centering
\begin{tabular}{ C{1cm} | C{1.5cm}  C{1.5cm}  C{1.5cm}  C{1.5cm}  C{1.5cm}   C{1.5cm}    C{1.5cm}   C{1.5cm}    }
\hline\hline
$L$ & 9 & 17 &  33 & 65 & 129 &  257 & 513 & 1025 \\
\hline
Q19 & -1.0512 & -1.2780 & -1.6309 & -2.1061 & -2.6404 & -3.2085 & -3.7936 & -4.3873\\
Q27 & -1.0575 & -1.2899 & -1.6491 & -2.1284 & -2.6842 & -3.3012 & -3.9555 & -4.6511\\
\hline\hline
\end{tabular}
\caption{Hartmann flow simulation: 10-based logarithm of the relative error achieved by the D3Q19-CM-LBM and the D3Q27-CM-LBM at $\mathrm{Ha=20}$.}
\label{TableHartmann}
\end{table*}

\subsection{Static bubble}
Now, the accuracy of the present approach is tested in the context of a multiphase flow simulated through the popular Shan-Chen model~\cite{shan1993lattice}. By introducing the so-called pseudo-potential $\psi = 1-\mathrm{exp}\left({-\rho} \right)$,  an interaction force
\begin{equation}
\bm{F} (\bm{x},t) = -G \psi(\bm{x},t) \sum_i w_i \bm{c}_i \psi(\bm{x}+\bm{c}_i,t),
\end{equation}
is used to mimic the molecular interactions leading to phase segregation, where $G$ is a parameter controlling the strength of the interaction. Let us consider a periodic domain consisting of $100$ lattice points in each direction. A droplet of variable radius $R \in [15:30]$ and density equal to $1.95$ is placed in the center of the domain, while the density is set to $0.15$ elsewhere. The kinematic viscosity is $\nu=0.0333$. The parameter $G$ is set equal to -5. 

\begin{figure}[tbp!]
\centering
\includegraphics[scale=1]{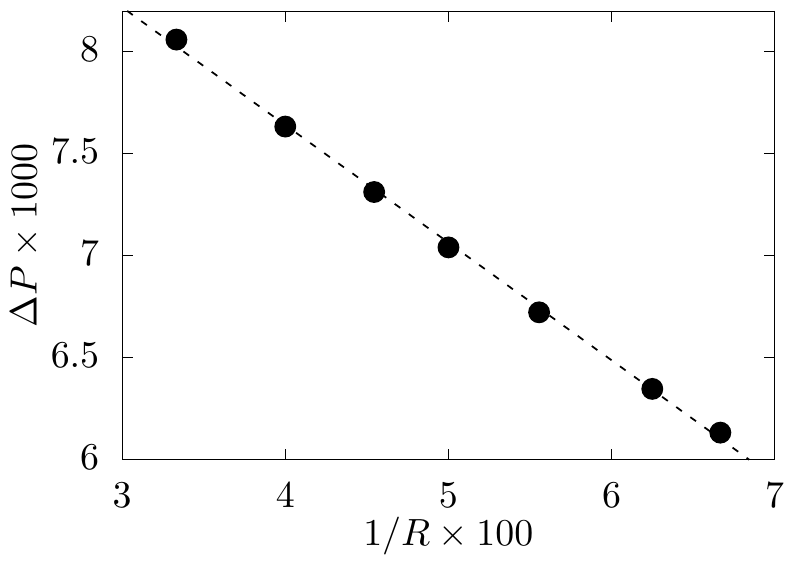}
\caption{Static bubble: pressure jump vs the inverse of the bubble radius. Data are fitted by a dashed black line.}
\label{Figure11}
\end{figure}

In Figure~\ref{Figure11}, the pressure jump across the interface is plotted against the inverse of the bubble radius. 
The linear evolution of the pressure jump with respect to the bubble radius proves that the present approach is able to account for surface tension as described by Young-Laplace's law.

\begin{figure*}[tbp!]
\centering
\includegraphics[width=0.75\textwidth]{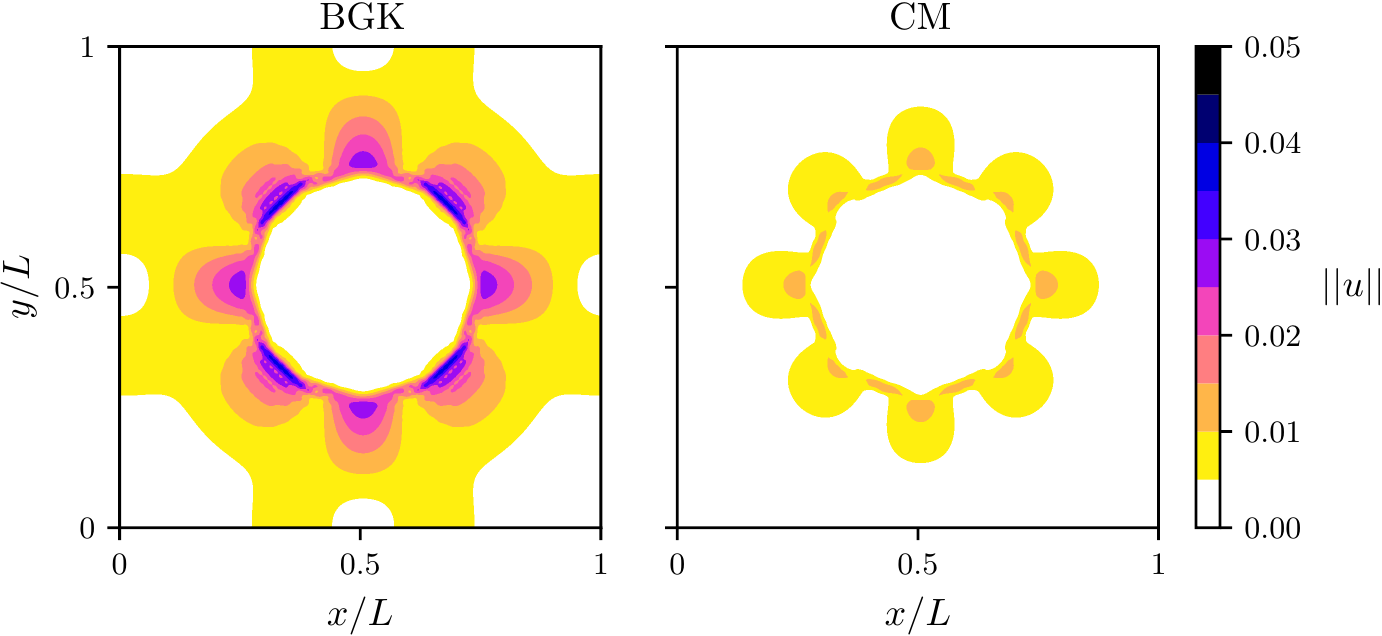}
\caption{Static bubble: map of the spurious currents by adopting two collision operators for the D3Q19-LBM (grid mesh composed of $L=100$ points in each direction).}
\label{Figure12}
\end{figure*}

Furthermore, we use this test to evaluate the ability of the present model to tackle spurious currents at the interface, that are well-known numerical artifacts affecting the Shan-Chen model. By considering $R=20$, this quantity is sketched in Figure~\ref{Figure12} together with findings from simulations carried out by using the D3Q19-BGK-LBM with extended equilibrium (\ref{eq:EqQ19RM}). The simplest collision model clearly exhibits strong anisotropic artifacts that undermine the stability and the accuracy of the run, and which are most likely related to poor spectral properties (dissipation and dispersion). Interestingly, this is drastically alleviated by the adoption of CMs with equilibration of CMs related to high-order kinetics and bulk viscosity. In fact, the resultant velocity map is considerably smoother and it shows a reduced magnitude. This might be related to the introduction of numerical (hyper-)viscosity induced by the equilibration of high-order moments~\cite{HOSSEINI_PRE_99_2019b,WISSOCQ_ARXIV_2020_07353} as well as the increased bulk viscosity~\cite{DELLAR_PRE_64_2001}.

\subsection{Rayleigh-Taylor instability}

\begin{figure*}
\centering
\includegraphics[width=0.99\textwidth]{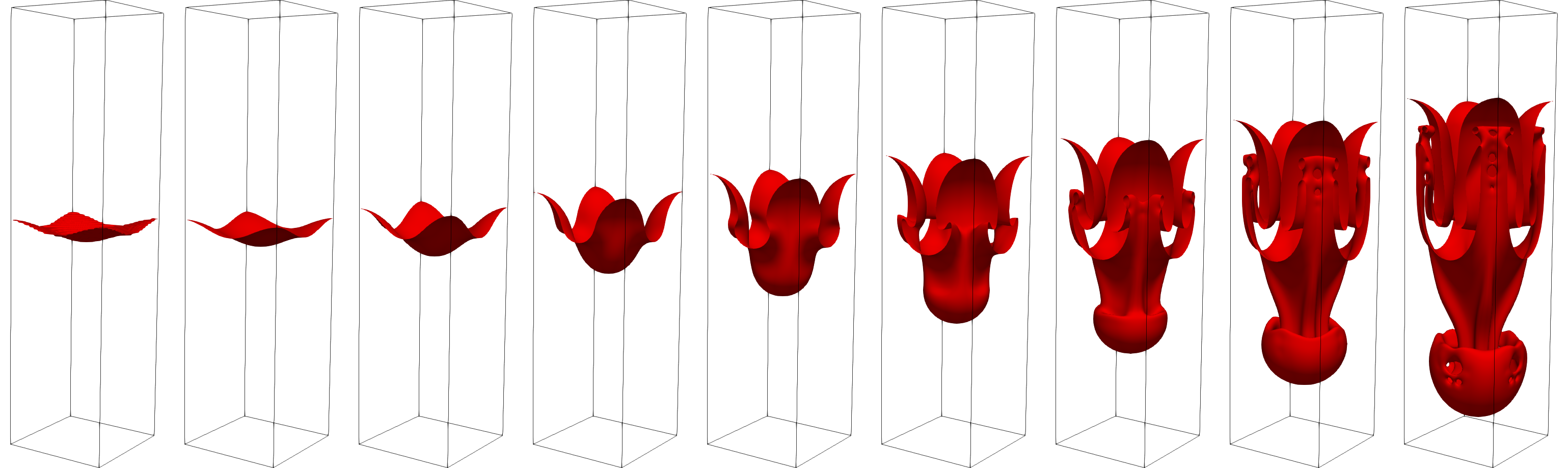}
\caption{Rayleigh-Taylor instability: time evolution of the interface at salient time instant, i.e., $t/t_0=0.0$, 0.5, 1.0, 1.5, 2.0, 2.5, 3.0, 3.5 and 4.0 (from left to right).}
\label{Figure13}
\end{figure*}

\begin{table*}[!htbp]
\centering
\begin{tabular}{ C{1cm} | C{2.5cm}  C{2.8cm}  C{2.8cm}  C{2.cm}  C{3cm}   C{1.cm}    }
\hline\hline
$t / t_0$ & D3Q19-CGM-CM & D3Q27-CGM-CM~\cite{DEROSIS_PoF_31_2019} &  D3Q27-CGM-MRT~\cite{saito2017lattice} & D3Q15-BGK~\cite{he1999three} & D3Q19-phase-field-MRT~\cite{WANG2016340} &  NS-CH~\cite{LEE20131466} \\
\hline
0.0 & 1.897 & 1.897 & 1.895 & 1.887 & 1.888 & 1.904\\
0.5 & 1.897 & 1.897 & 1.864 & 1.839 & 1.860 & 1.869\\
1.0 & 1.753 & 1.753 & 1.763 & 1.744 & 1.755 & 1.776\\
1.5 & 1.592 & 1.591 & 1.587 & 1.555 & 1.569 & 1.618\\
2.0 & 1.381 & 1.378 & 1.357 & 1.312 & 1.325 & 1.396\\
2.5 & 1.126 & 1.121 & 1.085 & 1.022 & 1.037 & 1.149\\
3.0 & 0.844 & 0.791 & 0.788 & 0.712 & 0.740 & 0.863\\
3.5 & 0.546 & 0.537 & 0.481 & 0.390 & 0.419 & 0.572\\
4.0 & 0.242 & 0.233 & 0.160 & 0.060 & 0.090 & 0.271\\
\hline\hline
\end{tabular}
\caption{Rayleigh-Taylor instability: time evolution of the position of the spike of the interface at salient time instants. Present results are compared to those from (i) the D3Q27-CGM-CM-LBM in \cite{DEROSIS_PoF_31_2019}, (ii) a D3Q27-CGM-MRT LB study based on the color-gradient method (CGM)~\cite{saito2017lattice}, (iii) a D3Q15-BGK LB model for multiphase flows~\cite{he1999three}, (iv) a D3Q19-phase-field-MRT LB scheme~\cite{WANG2016340}, and (v) a solution of the coupled Navier-Stokes-Cahn-Hilliard (NS-CH) equations~\cite{LEE20131466}. Reproduced from A. De Rosis, R. Huang, and C. Coreixas, Universal formulation of central-moments-based lattice Boltzmann method with external forcing for the simulation of multiphysics phenomena," Phys. Fluids 31, 117102 (2019), with the permission of AIP Publishing. Reproduced from X. He, R. Zhang, S. Chen, and G. D. Doolen, On the three-dimensional Rayleigh-Taylor instability," Phys. Fluids 11, 1 (1999), with the permission of AIP Publishing. Reproduced with permission from J. Comput. Sci. 17, (2016). Copyright 2016 Elsevier. Reproduced with permission from Comput. Math. Appl. Comput. Sci. 66, (2013). Copyright 2013 Elsevier.}
\label{Table6}
\end{table*}

We conclude the numerical campaign by simulating the Rayleigh-Taylor instability mechanism with a D3Q19-CM implementation of the color-gradient method (CGM)~\cite{PhysRevE.71.056702, Reis_2007, LECLAIRE20122237}. The interested reader can refer to App.~\ref{app:AppCG} for further details. Let us consider a three-dimensional domain of size $W \times 4W \times W$, with $W=64$, where a fluid of density {\color{black}$\rho_h=3$} is placed over a lighter one of density $\rho_l=1$. The fluid is initially at rest and initial conditions in terms of density read as follows
\begin{eqnarray}
\rho(\bm{x},0) &=& \rho_h, \, \mathrm{if} \, y>2W + 0.05W \left[ \cos \left( 2 \pi x \right)+\cos \left( 2 \pi z \right)  \right], \nonumber \\
\rho(\bm{x},0) &=& \rho_l, \, \mathrm{otherwise.}
\end{eqnarray}
The domain is periodic at every side, except for the top and bottom sections where the no-slip boundary condition is assigned. The flow is driven by a gravitational body force, that is
\begin{equation}
\bm{F} = - \left[ \rho(\bm{x},t) - \frac{\rho_h+\rho_l}{2}  \right] \bm{g},
\end{equation}
with $\bm{g}=(0, \, -g, \, 0)$, and $g$ chosen so that $t_0 = \sqrt{gW} = 0.04$~\cite{PhysRevE.71.056702}. The problem is governed by two dimensionless parameters, that are the Reynolds number 
$\displaystyle \mathrm{Re} = W\sqrt{gW}/\nu = 512$, and Atwood number, $\displaystyle \mathrm{At} = (\rho_h-\rho_l)/(\rho_h+\rho_l) = 0.5$.

In Figure~\ref{Figure13}, the evolution of the interface between the two fluids is sketched at salient time instants. {\color{black}Notice that the interface is identified as the set of lattice points where $[\rho_l(\bm{x})-\rho_h(\bm{x})]/[\rho_l(\bm{x})+\rho_h(\bm{x})]=0$.} A quantitative analysis of the results is reported in Table~\ref{Table6}. Present findings are compared to several models to assess its accuracy: (i) the D3Q27-CGM-CM-LBM recently proposed in Ref.~\cite{DEROSIS_PoF_31_2019}, (ii) a  D3Q27-CGM-MRT study~\cite{saito2017lattice}, (iii) a D3Q15-BGK LB model for multiphase flows~\cite{he1999three}, (iv) a D3Q19-phase-field MRT LB scheme~\cite{WANG2016340} and (v) a solution of the coupled Navier-Stokes-Cahn-Hilliard equations~\cite{LEE20131466}. From this, the present method shows a pretty good agreement with data from the literature, even though some discrepancies are also observed as $t/t_0$ grows, the latter being related to the equilibration of high-order moments as well as the increased bulk viscosity of both CMs-based algorithms. In any case, the reduction of the number of discrete velocities does not deteriorate the accuracy of the CM-LBM, and this confirms the good numerical properties of the proposed approach, as well as, its universality.

\section{Conclusions}
\label{SEC:IV}

In this paper, a three-dimensional lattice Boltzmann method has been proposed for the simulation of multiphysics phenomena (single phase, multiphase and magnetohydrodynamic flows). By adopting the D3Q19 velocity discretization instead of the more standard --but also more computationally intensive-- D3Q27 formulation, non-negligible gains are obtained in terms of both wall-clock time and memory consumption. 
In order to understand the limitations of such a choice, a large number of validation testcases were considered with a special focus on accuracy and stability discrepancies that would emerge from the adoption of the D3Q19 lattice.
Most of them confirmed that the latter discretization is pretty similar to its D3Q27 counterpart in terms of accuracy convergence and stability.
In fact, it is only for finite Mach numbers ( $\mathrm{Ma}\geq 0.2$) and under-resolved conditions that one should not reduce the number of discrete velocities in order not to face stability and accuracy issues. Apart from that, it seems quite safe to adopt the D3Q19 formulation in order to speed up the simulation of multiphysics flows.\\
\indent In parallel, we proved that by including up to fourth-order velocity terms in the equilibrium (as it is naturally the case for D3Q19-CM-LBMs), one can improve the stability of LBMs for the simulation of single phase flows in the low viscosity regime, and at moderate Mach numbers. One may then wonder if such a result can be extended to more complex flows. Corresponding investigations will be presented elsewhere.\\
\indent Eventually, as a possible extension to this work, one could include correction terms for velocity-dependent errors (those that are still present in the viscous stress tensor) in order to further improve the accuracy and stability of the present approach. This is motivated by the fact that these errors terms are usually non-negligible in under-resolved conditions (since they are proportional to the space step) and for finite values of the Mach number (because they depend on the velocity field). Hence, corrections terms employed in the context of compressible LBMs~\cite{FENG_JCP_394_2019,HOSSEINI_RSTA_378_2020,RENARD_ARXIV_2020_03644,RENARD_ARXIV_2020_08477} might further improve the present model for moderate Mach number flow simulations. 

\section*{Supplementary Material}
The script D3Q19CentralMoments.m allows us to perform all the symbolic manipulations to re-build the model proposed in this paper. 
The script D3Q27CentralMoments.m allows us to derive the central-moments-based scheme in the 27-velocities lattice discretization.

\begin{acknowledgments}
A.D.R. would like to thank Dr. T. Reis and Dr. S. Mohammed for valuable suggestions related to the setup of the simulations of the dipole-wall collisions.
\end{acknowledgments}

\section*{Data availability statement}
The data that supports the findings of this study are available within the article [and its supplementary material].
\appendix 

\section{Impact of the extended equilibrium \label{app:AppExtendedEquilibrium}}

\begin{figure*}[htbp!]
\centering
\subfigure{\includegraphics[width=0.9\textwidth]{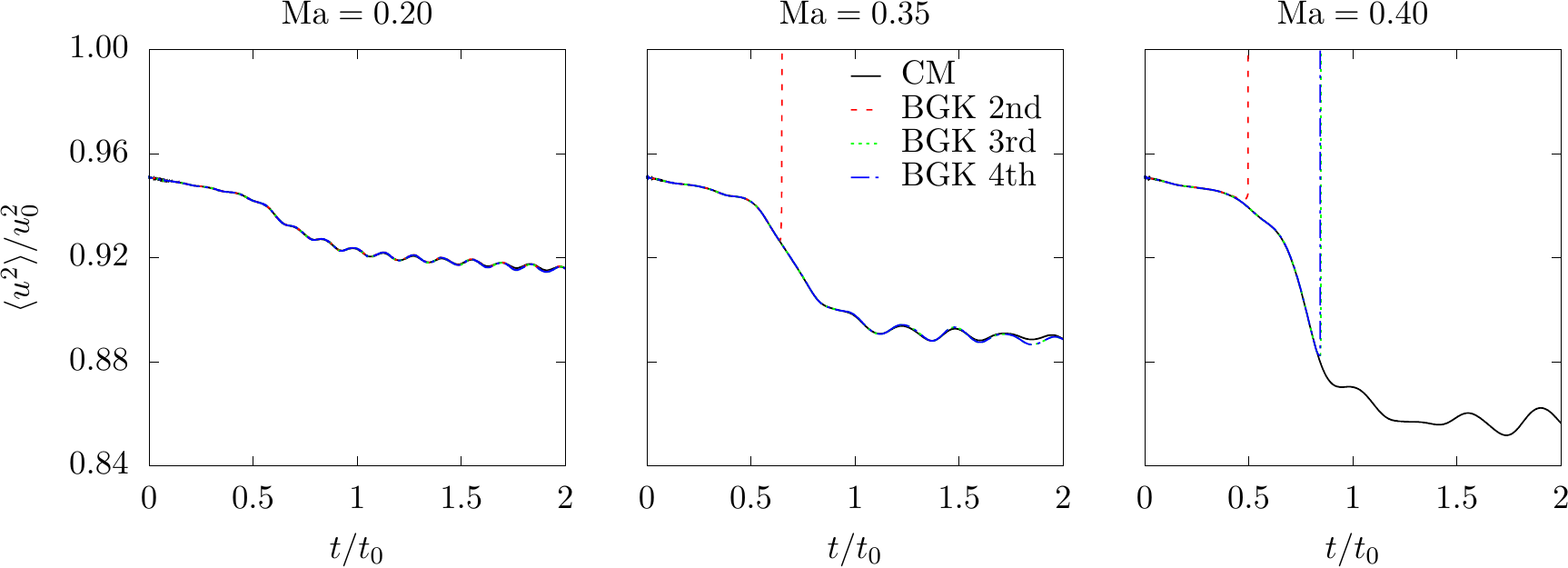}}
\caption{Double shear layer: evolution of the normalized kinetic energy, at different Mach numbers, for the D3Q19-CM-LBM (black continuous lines), the BGK with second-order equilibrium (red dashed line), the BGK with third-order equilibrium (green dotted line), and the BGK with fourth-order equilibrium (blue dash-dotted line).}
\label{Figure14}
\end{figure*}

The test in Sec.~\ref{DoubleShearLayer} is here adopted to demonstrate the impact of the extended equilibrium. In Figure~\ref{Figure14}, the time evolution of the normalized kinetic energy is plotted for different values of the Mach number, $\mathrm{Ma}=0.2, \, 0.35, \, 0.4$ by using four different approaches: (i) the present D3Q19-CM-LBM, the BGK-LBM with velocity terms up to the (ii) second order, (iii) third order, and (iv) fourth order. For the lowest value of Ma, all the runs can successfully simulate the whole desired time span and they produce similar results. More intriguing results are achieved as Ma grows. In fact, the poorer BGK-LBM blows up at $t/t_0 \sim 0.65$ when $\mathrm{Ma}=0.35$, while the BGK-LBMs with higher velocity terms and the D3Q19-CM-LBM are still stable. Therefore, we can assess that the introduction of higher-order velocity terms in the equilibrium population leads to an increase of the stability of the single-relaxation-time LBM. Then, when $\mathrm{Ma}=0.4$ all the BGK-LBMs fail, with the BGK-LBM with second-order equilibrium undergoes instabilities even before ($t/t_0 \sim 0.5$). The adoption of a more sophisticated equilibrium slightly alleviates this problem, as it is able to reach $t/t_0 \sim 0.86$ before blowing up. In the latter case, both third- and fourth-order equilibria lead to almost identical results.\\
\indent In the end, the extended equilibrium seems to be one of the stabilizating mechanism of the present LBM --in addition to the equilibration of high-order and bulk viscosity related moments. This is in accordance with results obtained through linear stability analyses and numerical simulations for various collision models in the context of D2Q9-LBMs~\cite{COREIXAS_PhD_2018,HOSSEINI_PRE_99_2019b,HOSSEINI_PRE_100_2019,COREIXAS_RSTA_378_2020}.

\section{Galilean invariant forcing scheme \label{app:AppForcingQ19}}
While the derivation of Galilean invariant forcing terms (for any kind of moment space) is rather straightforward in the context of multiphysics flow simulations based on the D3Q27 velocity discretization~\cite{DEROSIS_PoF_31_2019}, it is slightly more complex for the D3Q19 lattice since the latter is not built through tensor products of D1Q3 lattices in each direction $x$, $y$ and $z$. Nonetheless, by relying on the formulation based on the raw moment (RM) space, a general strategy can be proposed to construct D3Q19 formulations thanks to their D3Q27 counterparts~\cite{COREIXAS_PRE_100_2019}. The question is then: how can we move from the Gauss-Hermite formulation --that was proposed for the D3Q27 lattice in our previous study~\cite{DEROSIS_PoF_31_2019}-- to the RM formulation of interest? The naive manner would be to rewrite it in terms of Hermite moments (HMs) by replacing weights by their values (since this is a non-weighted formulation), and further converting it to a RM formulation using relationships between HMs and RMs. Nevertheless, there is a more straightforward way to achieve the same goal thanks to the moment matching approach. 
\begin{widetext}
Indeed, assuming that the form of the forcing RMs ($R^{\mathrm{RM}}_{pqr}$) is known, 
then the moment matching condition in the velocity space reads, $\forall p,q,r\leq2$
\begin{equation}
 \sum_i F^{\mathrm{RM}}_i c_{ix}^p c_{iy}^q c_{iz}^r = R^{\mathrm{RM}}_{pqr} 
\end{equation} 
which leads for the D3Q27 lattice to
\begin{align}
F^{\mathrm{RM,Q27}}_{(0, 0, 0)}&= R^{\mathrm{RM}}_{000}-R^{\mathrm{RM}}_{200}-R^{\mathrm{RM}}_{020}-R^{\mathrm{RM}}_{002}+R^{\mathrm{RM}}_{220}+R^{\mathrm{RM}}_{202}+R^{\mathrm{RM}}_{022}-R^{\mathrm{RM}}_{222}, \nonumber \\
F^{\mathrm{RM,Q27}}_{(\sigma, 0, 0)}&= \dfrac{1}{2} [ \sigma  R^{\mathrm{RM}}_{100} +R^{\mathrm{RM}}_{200} -\sigma  R^{\mathrm{RM}}_{120} -\sigma  R^{\mathrm{RM}}_{102} -R^{\mathrm{RM}}_{220} -R^{\mathrm{RM}}_{202} +\sigma  R^{\mathrm{RM}}_{122} +R^{\mathrm{RM}}_{222}],\nonumber \\
F^{\mathrm{RM,Q27}}_{(0, \lambda, 0)}&= \dfrac{1}{2} [\lambda  R^{\mathrm{RM}}_{010} +R^{\mathrm{RM}}_{020} -\lambda  R^{\mathrm{RM}}_{210} -\lambda  R^{\mathrm{RM}}_{012} -R^{\mathrm{RM}}_{220} -R^{\mathrm{RM}}_{022} +\lambda  R^{\mathrm{RM}}_{212} +R^{\mathrm{RM}}_{222}],\nonumber \\
F^{\mathrm{RM,Q27}}_{(0, 0, \chi)}&= \dfrac{1}{2} [  \chi  R^{\mathrm{RM}}_{001} +R^{\mathrm{RM}}_{002} -\chi  R^{\mathrm{RM}}_{201} -\chi  R^{\mathrm{RM}}_{021} -R^{\mathrm{RM}}_{202} -R^{\mathrm{RM}}_{022} +\chi  R^{\mathrm{RM}}_{221} +R^{\mathrm{RM}}_{222} ],\nonumber \\
F^{\mathrm{RM,Q27}}_{(\sigma, \lambda, 0)}&= \dfrac{1}{4} [ \sigma\lambda  R^{\mathrm{RM}}_{110}+\lambda  R^{\mathrm{RM}}_{210}+\sigma  R^{\mathrm{RM}}_{120}+R^{\mathrm{RM}}_{220}-\sigma\lambda    R^{\mathrm{RM}}_{112}-\lambda  R^{\mathrm{RM}}_{212}-\sigma  R^{\mathrm{RM}}_{122}-R^{\mathrm{RM}}_{222}  ],\nonumber \\
F^{\mathrm{RM,Q27}}_{(\sigma, 0, \chi)}&= \dfrac{1}{4} [\sigma\chi  R^{\mathrm{RM}}_{101} +\chi  R^{\mathrm{RM}}_{201} +\sigma  R^{\mathrm{RM}}_{102} +R^{\mathrm{RM}}_{202}-\sigma \chi R^{\mathrm{RM}}_{121} -\chi  R^{\mathrm{RM}}_{221} -\sigma  R^{\mathrm{RM}}_{122} -R^{\mathrm{RM}}_{222}   ],\nonumber \\
F^{\mathrm{RM,Q27}}_{(0, \lambda, \chi)}&= \dfrac{1}{4} [\lambda\chi  R^{\mathrm{RM}}_{011} +\chi  R^{\mathrm{RM}}_{021} +\lambda  R^{\mathrm{RM}}_{012} +R^{\mathrm{RM}}_{022} -\lambda\chi  R^{\mathrm{RM}}_{211}  -\chi  R^{\mathrm{RM}}_{221} -\lambda  R^{\mathrm{RM}}_{212} -R^{\mathrm{RM}}_{222}   ],\nonumber \\
F^{\mathrm{RM,Q27}}_{(\sigma, \lambda, \chi)}&= \dfrac{1}{8} [\sigma\lambda\chi R^{\mathrm{RM}}_{111}+ \lambda\chi  R^{\mathrm{RM}}_{211}+\sigma\chi  R^{\mathrm{RM}}_{121}+\sigma\lambda  R^{\mathrm{RM}}_{112}+\chi  R^{\mathrm{RM}}_{221}+\lambda  R^{\mathrm{RM}}_{212}+\sigma  R^{\mathrm{RM}}_{122}+R^{\mathrm{RM}}_{222}],\label{eq:FRMQ27}
\end{align}
where, for the sake of compactness, the tensor product notation has been adopted 
with $(\sigma,\lambda,\chi)\in\{\pm 1\}^3$. By discarding discrete velocities $(\pm1, \pm1, \pm1)$, the following constraints are obtained 
\begin{equation}
R^{\mathrm{RM}}_{111}=R^{\mathrm{RM}}_{211}=R^{\mathrm{RM}}_{121}=R^{\mathrm{RM}}_{112}=R^{\mathrm{RM}}_{221}=R^{\mathrm{RM}}_{212}=R^{\mathrm{RM}}_{122}=R^{\mathrm{RM}}_{222}=0,
\end{equation}
because $c_{ix}^p c_{iy}^q c_{iz}^r=0$ for these RMs. Consequently, the D3Q19 formulation of the forcing term then reads
\begin{align}
F^{\mathrm{RM,Q19}}_{0} &= R^{\mathrm{RM}}_{000} - R^{\mathrm{RM}}_{200} - R^{\mathrm{RM}}_{020} - R^{\mathrm{RM}}_{002} + R^{\mathrm{RM}}_{220} + R^{\mathrm{RM}}_{202} + R^{\mathrm{RM}}_{022},\nonumber \\
F^{\mathrm{RM,Q19}}_{1} &= \tfrac{1}{2} [\phantom{-} R^{\mathrm{RM}}_{100} + R^{\mathrm{RM}}_{200} - R^{\mathrm{RM}}_{120} - R^{\mathrm{RM}}_{102} - R^{\mathrm{RM}}_{220} - R^{\mathrm{RM}}_{202}],\nonumber \\
F^{\mathrm{RM,Q19}}_{2} &= \tfrac{1}{2} [\phantom{-} R^{\mathrm{RM}}_{010} + R^{\mathrm{RM}}_{020} - R^{\mathrm{RM}}_{210} - R^{\mathrm{RM}}_{012} - R^{\mathrm{RM}}_{220} - R^{\mathrm{RM}}_{022}],\nonumber \\
F^{\mathrm{RM,Q19}}_{3} &= \tfrac{1}{2} [\phantom{-} R^{\mathrm{RM}}_{001} + R^{\mathrm{RM}}_{002} - R^{\mathrm{RM}}_{201} - R^{\mathrm{RM}}_{021} - R^{\mathrm{RM}}_{202} - R^{\mathrm{RM}}_{022}],\nonumber \\
F^{\mathrm{RM,Q19}}_{4} &= \tfrac{1}{4} [\phantom{-} R^{\mathrm{RM}}_{110} - R^{\mathrm{RM}}_{210} - R^{\mathrm{RM}}_{120} + R^{\mathrm{RM}}_{220}],\nonumber \\
F^{\mathrm{RM,Q19}}_{5} &= \tfrac{1}{4} [\phantom{-} R^{\mathrm{RM}}_{110} + R^{\mathrm{RM}}_{210} + R^{\mathrm{RM}}_{120} + R^{\mathrm{RM}}_{220}],\nonumber \\
F^{\mathrm{RM,Q19}}_{6} &= \tfrac{1}{4} [\phantom{-} R^{\mathrm{RM}}_{101} - R^{\mathrm{RM}}_{201} - R^{\mathrm{RM}}_{102} + R^{\mathrm{RM}}_{202}],\nonumber \\
F^{\mathrm{RM,Q19}}_{7} &= \tfrac{1}{4} [\phantom{-} R^{\mathrm{RM}}_{101} + R^{\mathrm{RM}}_{201} + R^{\mathrm{RM}}_{102} + R^{\mathrm{RM}}_{202}],\nonumber \\
F^{\mathrm{RM,Q19}}_{8} &= \tfrac{1}{4} [\phantom{-} R^{\mathrm{RM}}_{011} - R^{\mathrm{RM}}_{021} - R^{\mathrm{RM}}_{012} + R^{\mathrm{RM}}_{022}],\nonumber \\
F^{\mathrm{RM,Q19}}_{9} &= \tfrac{1}{4} [\phantom{-} R^{\mathrm{RM}}_{011} + R^{\mathrm{RM}}_{021} + R^{\mathrm{RM}}_{012} + R^{\mathrm{RM}}_{022}],\nonumber \\
F^{\mathrm{RM,Q19}}_{10}&= \tfrac{1}{2} [- R^{\mathrm{RM}}_{100} + R^{\mathrm{RM}}_{200} + R^{\mathrm{RM}}_{120} + R^{\mathrm{RM}}_{102} - R^{\mathrm{RM}}_{220} - R^{\mathrm{RM}}_{202}],\nonumber \\
F^{\mathrm{RM,Q19}}_{11}&= \tfrac{1}{2} [- R^{\mathrm{RM}}_{010} + R^{\mathrm{RM}}_{020} + R^{\mathrm{RM}}_{210} + R^{\mathrm{RM}}_{012} - R^{\mathrm{RM}}_{220} - R^{\mathrm{RM}}_{022}],\nonumber \\
F^{\mathrm{RM,Q19}}_{12}&= \tfrac{1}{2} [- R^{\mathrm{RM}}_{001} + R^{\mathrm{RM}}_{002} + R^{\mathrm{RM}}_{201} + R^{\mathrm{RM}}_{021} - R^{\mathrm{RM}}_{202} - R^{\mathrm{RM}}_{022}],\nonumber \\
F^{\mathrm{RM,Q19}}_{13}&= \tfrac{1}{4} [- R^{\mathrm{RM}}_{110} + R^{\mathrm{RM}}_{210} - R^{\mathrm{RM}}_{120} + R^{\mathrm{RM}}_{220}],\nonumber \\
F^{\mathrm{RM,Q19}}_{14}&= \tfrac{1}{4} [- R^{\mathrm{RM}}_{110} - R^{\mathrm{RM}}_{210} + R^{\mathrm{RM}}_{120} + R^{\mathrm{RM}}_{220}],\nonumber \\
F^{\mathrm{RM,Q19}}_{15}&= \tfrac{1}{4} [- R^{\mathrm{RM}}_{101} + R^{\mathrm{RM}}_{201} - R^{\mathrm{RM}}_{102} + R^{\mathrm{RM}}_{202}],\nonumber \\
F^{\mathrm{RM,Q19}}_{16}&= \tfrac{1}{4} [- R^{\mathrm{RM}}_{101} - R^{\mathrm{RM}}_{201} + R^{\mathrm{RM}}_{102} + R^{\mathrm{RM}}_{202}],\nonumber \\
F^{\mathrm{RM,Q19}}_{17}&= \tfrac{1}{4} [- R^{\mathrm{RM}}_{011} + R^{\mathrm{RM}}_{021} - R^{\mathrm{RM}}_{012} + R^{\mathrm{RM}}_{022}],\nonumber \\
F^{\mathrm{RM,Q19}}_{18}&= \tfrac{1}{4} [- R^{\mathrm{RM}}_{011} - R^{\mathrm{RM}}_{021} + R^{\mathrm{RM}}_{012} + R^{\mathrm{RM}}_{022}],\nonumber \\
F^{\mathrm{RM,Q19}}_{19\ldots 26}&= 0, \label{eq:FRMQ19}
\end{align}
\end{widetext}
where forcing terms corresponding to discrete velocities $(\pm1, \pm1, \pm1)$, which are $F^{\mathrm{RM,Q19}}_{20\ldots 26}$, naturally vanish in the context of RMs. The only remaining task is to actually determine what are the formulas that should be used for $R^{\mathrm{RM}}_{pqr}$. One can start from $R^{\mathrm{HM}}_{pqr}$ that were derived in our previous work~\cite{DEROSIS_PoF_31_2019}
\begin{align}
R^{\mathrm{HM}}_{000}&=0,\nonumber \\
R^{\mathrm{HM}}_{100}&=F_x,\nonumber \\
R^{\mathrm{HM}}_{010}&=F_y,\nonumber \\
R^{\mathrm{HM}}_{001}&=F_z,\nonumber \\
R^{\mathrm{HM}}_{200}&=2 u_x F_x,\nonumber \\
R^{\mathrm{HM}}_{020}&=2 u_y F_y,\nonumber \\
R^{\mathrm{HM}}_{002}&=2 u_z F_z,\nonumber \\
R^{\mathrm{HM}}_{110}&=u_x F_y + u_y F_x,\nonumber \\
R^{\mathrm{HM}}_{101}&=u_x F_z + u_z F_x,\nonumber \\
R^{\mathrm{HM}}_{011}&=u_y F_z + u_z F_y,\nonumber \\
R^{\mathrm{HM}}_{210}&=u_x(u_x F_y + 2u_y F_x),\nonumber \\
R^{\mathrm{HM}}_{120}&=u_y(u_y F_x + 2u_x F_y),\nonumber \\
R^{\mathrm{HM}}_{201}&=u_x(u_x F_z + 2u_z F_x),\nonumber \\
R^{\mathrm{HM}}_{102}&=u_z(u_z F_x + 2u_x F_z),\nonumber \\
R^{\mathrm{HM}}_{021}&=u_y(u_y F_z + 2u_z F_y),\nonumber \\
R^{\mathrm{HM}}_{012}&=u_z(u_z F_y + 2u_y F_z),\nonumber \\
R^{\mathrm{HM}}_{220}&=2 u_x u_y(u_x F_y + u_y F_x),\nonumber \\
R^{\mathrm{HM}}_{202}&=2 u_x u_z(u_x F_z + u_z F_x),\nonumber \\
R^{\mathrm{HM}}_{022}&=2 u_y u_z(u_y F_z + u_z F_y),\label{eq:HMQ19}
\end{align}
and compute $R^{\mathrm{RM}}_{pqr}$ through relationships between HMs and RMs. Another way to do it is by starting from central moments (CMs) of the forcing term~\cite{FEI_PRE_97_2018}
\begin{align}
R^{\mathrm{CM}}_{000}&=0,\nonumber \\
R^{\mathrm{CM}}_{100}&=F_x,\nonumber \\
R^{\mathrm{CM}}_{010}&=F_y,\nonumber \\
R^{\mathrm{CM}}_{001}&=F_z,\nonumber \\
R^{\mathrm{CM}}_{200}&=0,\nonumber \\
R^{\mathrm{CM}}_{020}&=0,\nonumber \\
R^{\mathrm{CM}}_{002}&=0,\nonumber \\
R^{\mathrm{CM}}_{110}&=0,\nonumber \\
R^{\mathrm{CM}}_{101}&=0,\nonumber \\
R^{\mathrm{CM}}_{011}&=0,\nonumber \\
R^{\mathrm{CM}}_{210}&=c_s^2 F_y,\nonumber \\
R^{\mathrm{CM}}_{120}&=c_s^2 F_x,\nonumber \\
R^{\mathrm{CM}}_{201}&=c_s^2 F_z,\nonumber \\
R^{\mathrm{CM}}_{102}&=c_s^2 F_x,\nonumber \\
R^{\mathrm{CM}}_{021}&=c_s^2 F_z,\nonumber \\
R^{\mathrm{CM}}_{012}&=c_s^2 F_y,\nonumber \\
R^{\mathrm{CM}}_{220}&=0,\nonumber \\
R^{\mathrm{CM}}_{202}&=0,\nonumber \\
R^{\mathrm{CM}}_{022}&=0,\label{eq:CMQ19}
\end{align}
and compute $R^{\mathrm{RM}}_{pqr}$ through relationships between CMs and RMs. One can even start from the central Hermite moment (CHM) framework which is the only one leading to velocity- and lattice-independent moments of the forcing term, i.e.,
\begin{equation}
R^{\mathrm{CHM}}_{100}=F_x,\,
R^{\mathrm{CHM}}_{010}=F_y,\,
R^{\mathrm{CHM}}_{001}=F_z,
\label{eq:CHMQ19}
\end{equation}
and the other CHMs equal to zero, and eventually coming back to RMs through formulas provided in~\cite{COREIXAS_PRE_100_2019}. Whatever the approach considered, RMs of the forcing term eventually read
\begin{align}
R^{\mathrm{RM}}_{000}&=0,\nonumber \\
R^{\mathrm{RM}}_{100}&=F_x,\nonumber \\
R^{\mathrm{RM}}_{010}&=F_y,\nonumber \\
R^{\mathrm{RM}}_{001}&=F_z,\nonumber \\
R^{\mathrm{RM}}_{200}&=2 u_x F_x,\nonumber \\
R^{\mathrm{RM}}_{020}&=2 u_y F_y,\nonumber \\
R^{\mathrm{RM}}_{002}&=2 u_z F_z,\nonumber \\
R^{\mathrm{RM}}_{110}&=u_x F_y + u_y F_x,\nonumber \\
R^{\mathrm{RM}}_{101}&=u_x F_z + u_z F_x,\nonumber \\
R^{\mathrm{RM}}_{011}&=u_y F_z + u_z F_y,\nonumber \\
R^{\mathrm{RM}}_{210}&=c_s^2 F_y + u_x(u_x F_y + 2u_y F_x),\nonumber \\
R^{\mathrm{RM}}_{120}&=c_s^2 F_x + u_y(u_y F_x + 2u_x F_y),\nonumber \\
R^{\mathrm{RM}}_{201}&=c_s^2 F_z + u_x(u_x F_z + 2u_z F_x),\nonumber \\
R^{\mathrm{RM}}_{102}&=c_s^2 F_x + u_z(u_z F_x + 2u_x F_z),\nonumber \\
R^{\mathrm{RM}}_{021}&=c_s^2 F_z + u_y(u_y F_z + 2u_z F_y),\nonumber \\
R^{\mathrm{RM}}_{012}&=c_s^2 F_y + u_z(u_z F_y + 2u_y F_z),\nonumber \\
R^{\mathrm{RM}}_{220}&=2 [u_x u_y(u_x F_y + u_y F_x) + c_s^2 (u_x F_x + u_y F_y)],\nonumber \\
R^{\mathrm{RM}}_{202}&=2 [u_x u_z(u_x F_z + u_z F_x) + c_s^2 (u_x F_x + u_z F_z)],\nonumber \\
R^{\mathrm{RM}}_{022}&=2 [u_y u_z(u_y F_z + u_z F_y) + c_s^2 (u_y F_y + u_z F_z)].\label{eq:RMQ19}
\end{align}
Interestingly, due to relationships between all moment spaces, $R^{\mathrm{HM}}_{pqr}$ can be obtained from $R^{\mathrm{RM}}_{pqr}$ by neglecting lattice-dependent terms (those proportional to $c_s$), while $R^{\mathrm{CM}}_{pqr}$ is obtained by discarding velocity-dependent terms. Eventually, $R^{\mathrm{CHM}}_{pqr}$ is derived by neglecting both lattice- and velocity-depend terms in $R^{\mathrm{RM}}_{pqr}$. 

\begin{widetext}
\section{Raw moment formulation \label{app:AppRMQ19}}
Hereafter, the universal nature of our D3Q19-CM-LBM is highlighted by rewriting it in the RM space~\citep{Fei2017,Fei2018a}. The starting point is to notice that the matrix $\mathbf{T}$ can be written as the product of two contributions, i.e.
\begin{equation}
\mathbf{T} = \mathbf{N} \mathbf{M}.
\end{equation}
The transformation matrix $\mathbf{M}$ can be computed as
	\begin{equation}
	\mathbf{M} =
	\left [
\begin{array}{c}
\bra{|\boldsymbol{c}_i|^0} \\
\bra{ c_{ix}}\\ 
\bra{ c_{iy}} \\
\bra{ c_{iz}} \\  
\bra{ c_{ix}^2+ c_{iy}^2 + c_{iz}^2 } \\
\bra{ c_{ix}^2- c_{iy}^2 } \\ 
\bra{ c_{iy}^2- c_{iz}^2}\\
\bra{ c_{ix} c_{iy}} \\
\bra{ c_{ix} c_{iz} } \\ 
\bra{ c_{iy} c_{iz}} \\
\bra{c_{ix}^2c_{iy}}\\
\bra{c_{ix}c_{iy}^2} \\
\bra{ c_{ix}^2c_{iz}}\\
\bra{ c_{ix}c_{iz}^2} \\
\bra{c_{iy}^2c_{iz}} \\
\bra{ c_{iy}c_{iz}^2} \\
\bra{ c_{ix}^2c_{iy}^2} \\
\bra{ c_{ix}^2c_{iz}^2} \\
\bra{c_{iy}^2c_{iy}^2}
\end{array}
\right]  = 
\left [
\begin{array}{ccccccccccccccccccc}
     1&     1&    1&    1&    1&    1&    1&    1&    1&    1&    1&    1&    1&    1&    1&    1&    1&    1&    1\\
     0&    -1&    0&    0&   -1&   -1&   -1&   -1&    0&    0&    1&    0&    0&    1&    1&    1&    1&    0&    0\\
     0&    0&   -1&    0&   -1&    1&    0&    0&   -1&   -1&    0&    1&    0&    1&   -1&    0&    0&    1&    1\\
     0&     0&    0&   -1&    0&    0&   -1&    1&   -1&    1&    0&    0&    1&    0&    0&    1&   -1&    1&   -1\\
     0&     1&    1&    1&    2&    2&    2&    2&    2&    2&    1&    1&    1&    2&    2&    2&    2&    2&    2\\
     0&     1&   -1&    0&    0&    0&    1&    1&   -1&   -1&    1&   -1&    0&    0&    0&    1&    1&   -1&   -1\\
     0&     0&    1&   -1&    1&    1&   -1&   -1&    0&    0&    0&    1&   -1&    1&    1&   -1&   -1&    0&    0\\
     0&    0&    0&    0&    1&   -1&    0&    0&    0&    0&    0&    0&    0&    1&   -1&    0&    0&    0&    0\\
     0&    0&    0&    0&    0&    0&    1&   -1&    0&    0&    0&    0&    0&    0&    0&    1&   -1&    0&    0\\
     0&    0&    0&    0&    0&    0&    0&    0&    1&   -1&    0&    0&    0&    0&    0&    0&    0&    1&   -1\\
     0&    0&    0&    0&   -1&    1&    0&    0&    0&    0&    0&    0&    0&    1&   -1&    0&    0&    0&    0\\
     0&    0&    0&    0&   -1&   -1&    0&    0&    0&    0&    0&    0&    0&    1&    1&    0&    0&    0&    0\\
     0&    0&    0&    0&    0&    0&   -1&    1&    0&    0&    0&    0&    0&    0&    0&    1&   -1&    0&    0\\
     0&    0&    0&    0&    0&    0&   -1&   -1&    0&    0&    0&    0&    0&    0&    0&    1&    1&    0&    0\\
     0&    0&    0&    0&    0&    0&    0&    0&   -1&    1&    0&    0&    0&    0&    0&    0&    0&    1&   -1\\
     0&    0&    0&    0&    0&    0&    0&    0&   -1&   -1&    0&    0&    0&    0&    0&    0&    0&    1&    1\\
     0&    0&    0&    0&    1&    1&    0&    0&    0&    0&    0&    0&    0&    1&    1&    0&    0&    0&    0\\
     0&    0&    0&    0&    0&    0&    1&    1&    0&    0&    0&    0&    0&    0&    0&    1&    1&    0&    0\\
     0&    0&    0&    0&    0&    0&    0&    0&    1&    1&    0&    0&    0&    0&    0&    0&    0&    1&    1
\end{array}
\right] ,	
	\end{equation}	
Notably, it is obtained without performing the shifting of the lattice fluid directions by the local fluid velocity. Then, the shift matrix is $\mathbf{N} = \mathbf{T} \mathbf{M}^{-1}$.
The transformation matrix $\mathbf{M}$ transforms the distribution functions into the raw moments. The shift matrix $\mathbf{N}$ transforms the RMs into the CMs, and is a lower-triangular matrix. Notably, the shift matrix $\mathbf{N}$ was originally introduced by Asinari \cite{asinari2008generalized}. If this shift is neglected, i.e. $\mathbf{N}=\mathbf{I}$, then $\mathbf{T} = \mathbf{M}$. In other words, the classical (RMs-based) multiple-relaxation-time can be viewed as a particular case of a general (CMs-based) multiple-relaxation-time LBM. For practical implementation, it is easier to replace the above ``one-step" reconstruction by the ``two-step" reconstruction~\citep{Fei2017}. In short, Eq.~(\ref{system}) can be rewritten as
\begin{equation}
\ket{f_i^{\star}} 
= \mathbf{M}^{-1} \mathbf{N}^{-1} \ket{k_i^{\star} },
\end{equation}
where we first compute post-collision raw moments $\ket{r_i^{\star}} = \left[ r_0^{\star},\, \ldots, \, r_i^{\star},\, \ldots, \, r_{18}^{\star}   \right]^{\top}$ as
\begin{equation} \label{EqGetRawFromCentral}
\ket{r_i^{\star}} 					  =  {\mathbf{N}^{-1}} \ket{k_i^{\star} },
\end{equation}
that are equal to
\begin{eqnarray} \label{rawQ19}
r_0^{\star} &=& \rho,\nonumber \\
r_1^{\star} &=& \rho u_x  +k_1^{\star},\nonumber \\
r_2^{\star} &=& \rho u_y +k_2^{\star},\nonumber \\
r_3^{\star} &=& \rho u_z +k_3^{\star},\nonumber \\
r_4^{\star} &=& \rho (u_x^2+u_y^2+u_z^2) +2 u_x k_1^{\star}+2 u_y k_2^{\star}+2 u_z k_3^{\star}+k_4^{\star},\nonumber \\
r_5^{\star} &=& \rho (u_x^2-u_y^2) +2 u_x k_1^{\star}-2 u_y k_2^{\star}+k_5^{\star},\nonumber \\
r_6^{\star} &=& \rho (u_y^2-u_z^2) +2 u_y k_2^{\star}-2 u_z k_3^{\star}+k_6^{\star},\nonumber \\
r_7^{\star} &=& \rho u_x u_y +u_y k_1^{\star}+u_x k_2^{\star}+k_7^{\star},\nonumber \\
r_8^{\star} &=& \rho u_x u_z +u_z k_1^{\star}+u_x k_3^{\star}+k_8^{\star},\nonumber \\
r_9^{\star} &=& \rho u_y u_z +u_z k_2^{\star}+u_y k_3^{\star}+k_9^{\star},\nonumber \\
r_{10}^{\star} &=& \rho u_x^2 u_y +2 u_x u_y k_1^{\star}+u_x^2 k_2^{\star}+u_y k_4^{\star} c_s^2+2 u_y k_5^{\star} c_s^2+u_y k_6^{\star} c_s^2+2 u_x k_7^{\star}+k_{10}^{\star},\nonumber \\
r_{11}^{\star} &=& \rho u_x u_y^2 +u_y^2 k_1^{\star}+2 u_x u_y k_2^{\star}+u_x k_4^{\star} c_s^2-u_x k_5^{\star} c_s^2+u_x k_6^{\star} c_s^2+2 u_y k_7^{\star}+k_{11}^{\star},\nonumber \\
r_{12}^{\star} &=& \rho u_x^2 u_z +2 u_x u_z k_1^{\star}+u_x^2 k_3^{\star}+u_z k_4^{\star} c_s^2+2 u_z k_5^{\star} c_s^2+u_z k_6^{\star} c_s^2+2 u_x k_8^{\star}+k_{12}^{\star},\nonumber \\
r_{13}^{\star} &=& \rho u_x u_z^2 +u_z^2 k_1^{\star}+2 u_x u_z k_3^{\star}+u_x k_4^{\star} c_s^2-u_x k_5^{\star} c_s^2-2 u_x k_6^{\star} c_s^2+2 u_z k_8^{\star}+k_{13}^{\star},\nonumber \\
r_{14}^{\star} &=& \rho u_y^2 u_z +2 u_y u_z k_2^{\star}+u_y^2 k_3^{\star}+u_z k_4^{\star} c_s^2-u_z k_5^{\star} c_s^2+u_z k_6^{\star} c_s^2+2 u_y k_9^{\star}+k_{14}^{\star},\nonumber \\
r_{15}^{\star} &=& \rho u_y u_z^2 +u_z^2 k_2^{\star}+2 u_y u_z k_3^{\star}+u_y k_4^{\star} c_s^2-u_y k_5^{\star} c_s^2-2 u_y k_6^{\star} c_s^2+2 u_z k_9^{\star}+k_{15}^{\star},\nonumber \\
r_{16}^{\star} &=& \rho u_x^2 u_y^2 +2 u_x u_y^2 k_1^{\star}+2 u_x^2 u_y k_2^{\star}+(u_x^2+u_y^2) k_4^{\star} c_s^2+(2 u_y^2-u_x^2) k_5^{\star}+(u_x^2+u_y^2) k_6^{\star} c_s^2+4 u_x u_y k_7^{\star}+2 u_y k_{10}^{\star}+2 u_x k_{11}^{\star}+k_{16}^{\star},\nonumber \\
r_{17}^{\star} &=& \rho u_x^2 u_z^2 +2 u_x u_z^2 k_1^{\star}+2 u_x^2 u_z k_3^{\star}+(u_x^2+u_z^2) k_4^{\star} c_s^2+(2 u_z^2-u_x^2) k_5^{\star}+(u_z^2-2 u_x^2) k_6^{\star}c_s^2+4 u_x u_z k_8^{\star}+2 u_z k_{12}^{\star}+2 u_x k_{13}^{\star}+k_{17}^{\star},\nonumber \\
r_{18}^{\star} &=& \rho u_y^2 u_z^2 +2 u_y u_z^2 k_2^{\star}+2 u_y^2 u_z k_3^{\star}+(u_y^2+u_z^2) k_4^{\star} c_s^2+(-u_y^2-u_z^2) k_5^{\star} c_s^2+(u_z^2-2 u_y^2) k_6^{\star}+4 u_y u_z k_9^{\star}+2 u_z k_{14}^{\star}+2 u_y k_{15}^{\star}+k_{18}^{\star}.
\end{eqnarray}
and then we transform into populations
\begin{equation}
\ket{f_i^{\star}}  = \mathbf{M}^{-1} \ket{r_i^{\star}},
\end{equation}
i.e.
\begin{eqnarray} \label{pdfQ19}
f_0^{\star} &=&  r_0^{\star} - r_4^{\star} + r_{16}^{\star} + r_{17}^{\star} + r_{18}^{\star},\nonumber \\
f_1^{\star} &=&  (r_{11}^{\star}+r_{13}^{\star}-r_1^{\star}-r_{16}^{\star}-r_{17}^{\star})/2+(r_4^{\star}+2r_5^{\star}+r_6^{\star})c_s^2/2,\nonumber \\
f_2^{\star} &=&  (r_{10}^{\star}+r_{15}^{\star}-r_2^{\star}-r_{16}^{\star}-r_{18}^{\star})/2+(r_4^{\star}+r_6^{\star}-r_5^{\star})c_s^2/2,\nonumber \\
f_3^{\star} &=&  (r_{12}^{\star}+r_{14}^{\star}-r_3^{\star} -r_{17}^{\star}-r_{18}^{\star})/2+(r_4^{\star}-r_5^{\star}-2-r_6^{\star})c_s^2/2,\nonumber \\
f_4^{\star} &=&  (r_7^{\star}+r_{16}^{\star}-r_{10}^{\star}-r_{11}^{\star})/4,\nonumber \\
f_5^{\star} &=&  (r_{10}^{\star}+r_{16}^{\star}-r_7^{\star}-r_{11}^{\star})/4,\nonumber \\
f_6^{\star} &=&  (r_8^{\star}+r_{17}^{\star}-r_{12}^{\star}-r_{13}^{\star})/4,\nonumber \\
f_7^{\star} &=&  (r_{12}^{\star}+r_{17}^{\star}-r_8^{\star}-r_{13}^{\star})/4,\nonumber \\
f_8^{\star} &=&  (r_9^{\star}+r_{18}^{\star}-r_{14}^{\star}-r_{15}^{\star})/4,\nonumber \\
f_9^{\star} &=&  (r_{14}^{\star}+r_{18}^{\star}-r_9^{\star}-r_{15}^{\star})/4,\nonumber \\
f_{10}^{\star} &=&  (r_1^{\star}-r_{11}^{\star}-r_{13}^{\star}-r_{16}^{\star}-r_{17}^{\star})/2+(r_4^{\star}+2r_5^{\star}+r_6^{\star})c_s^2/2,\nonumber \\
f_{11}^{\star} &=&  (r_2^{\star}-r_{10}^{\star}-r_{15}^{\star}-r_{16}^{\star}-r_{18}^{\star})/2+(r_4^{\star}+r_6^{\star}-r_5^{\star})c_s^2/2,\nonumber \\
f_{12}^{\star} &=&  (r_3^{\star}-r_{12}^{\star}-r_{14}^{\star}-r_{17}^{\star}-r_{18}^{\star})/2+(r_4^{\star}-r_5^{\star}-r_6^{\star})c_s^2/2,\nonumber \\
f_{13}^{\star} &=&  (r_7^{\star}+r_{10}^{\star}+r_{11}^{\star}+r_{16}^{\star})/4,\nonumber \\
f_{14}^{\star} &=&  (r_{11}^{\star}+r_{16}^{\star}-r_7^{\star}-r_{10}^{\star})/4,\nonumber \\
f_{15}^{\star} &=&  (r_8^{\star}+r_{12}^{\star}+r_{13}^{\star}+r_{17}^{\star})/4,\nonumber \\
f_{16}^{\star} &=&  (r_{13}^{\star}+r_{17}^{\star}-r_8^{\star}-r_{12}^{\star})/4,\nonumber \\
f_{17}^{\star} &=&  (r_9^{\star}+r_{14}^{\star}+r_{15}^{\star}+r_{18}^{\star})/4,\nonumber \\
f_{18}^{\star} &=&  (r_{15}^{\star}+r_{18}^{\star}-r_9^{\star}-r_{14}^{\star})/4.
\end{eqnarray}
\end{widetext}
Finally, it should be noted that, when $\mathbf{N}=\mathbf{I}$, particular attention should be paid to the computation of pre-collision and equilibrium moments, as the \textit{un-shifted} transformation matrix $\mathbf{T} = \mathbf{M}$ is used. Indeed, equilibrium moments read as follows
\begin{eqnarray}
k_0^{eq} &=& \rho, \nonumber \\
k_1^{eq} &=& \rho u_x, \nonumber \\
k_2^{eq} &=& \rho u_y, \nonumber \\
k_3^{eq} &=& \rho u_z, \nonumber \\
k_4^{eq} &=& \rho \left(3c_s^2 +u_x^2+u_y^2+u_z^2 \right), \nonumber \\
k_5^{eq} &=& \rho \left( u_x^2-u_y^2  \right), \nonumber \\
k_6^{eq} &=& \rho \left( u_y^2-u_z^2  \right)\nonumber \\
k_7^{eq} &=& \rho u_x u_y , \nonumber \\
k_8^{eq} &=& \rho u_x u_z , \nonumber \\
k_9^{eq} &=& \rho u_y u_z , \nonumber \\
k_{10}^{eq} &=& \rho u_y \left(u_x^2+c_s^2 \right) , \nonumber \\
k_{11}^{eq} &=& \rho u_x \left(u_y^2+c_s^2  \right) , \nonumber \\
k_{12}^{eq} &=& \rho u_z \left(u_x^2+c_s^2 \right), \nonumber \\
k_{13}^{eq} &=& \rho u_x \left(u_z^2+c_s^2 \right), \nonumber \\
k_{14}^{eq} &=& \rho u_z \left(u_y^2+c_s^2 \right), \nonumber \\
k_{15}^{eq} &=& \rho u_y \left(u_z^2+c_s^2 \right), \nonumber \\
k_{16}^{eq} &=& \rho \left(u_x^2+c_s^2   \right)  \left(u_y^2+c_s^2   \right), \nonumber \\
k_{17}^{eq} &=& \rho \left(u_x^2+c_s^2   \right)  \left(u_z^2+c_s^2   \right), \nonumber \\
k_{18}^{eq} &=& \rho \left(u_y^2+c_s^2   \right)  \left(u_z^2+c_s^2   \right),
\end{eqnarray}
and the resultant post-collision state is
\begin{eqnarray}
k_0^{\star} &=& \rho, \nonumber \\
k_1^{\star} &=& F_x/2 + \rho u_x, \nonumber \\
k_2^{\star} &=& F_y/2 + \rho u_y, \nonumber \\
k_3^{\star} &=& F_z/2 + \rho u_z, \nonumber \\
k_4^{\star} &=& \rho \left(3 c_s^2 +u_x^2+u_y^2+u_z^2 \right), \nonumber \\
k_5^{\star} &=& \left(1-\omega \right)k_5 + \omega \rho \left( u_x^2-u_y^2  \right), \nonumber \\
k_6^{\star} &=& \left(1-\omega \right)k_6 + \omega \rho \left( u_y^2-u_z^2  \right)\nonumber \\
k_7^{\star} &=& \left(1-\omega \right)k_7 + \omega \rho u_x u_y , \nonumber \\
k_8^{\star} &=& \left(1-\omega \right)k_8 + \omega \rho u_x u_z , \nonumber \\
k_9^{\star} &=& \left(1-\omega \right)k_9 + \omega \rho u_y u_z , \nonumber \\
k_{10}^{\star} &=& F_y c_s^2 /2 + \rho u_y \left(u_x^2+c_s^2 \right) , \nonumber \\
k_{11}^{\star} &=& F_x c_s^2 /2 + \rho u_x \left(u_y^2+c_s^2  \right) , \nonumber \\
k_{12}^{\star} &=& F_z c_s^2 /2 + \rho u_z \left(u_x^2+c_s^2 \right), \nonumber \\
k_{13}^{\star} &=& F_x c_s^2 /2 + \rho u_x \left(u_z^2+c_s^2 \right), \nonumber \\
k_{14}^{\star} &=& F_z c_s^2 /2 + \rho u_z \left(u_y^2+c_s^2 \right), \nonumber \\
k_{15}^{\star} &=& F_y c_s^2 /2 + \rho u_y \left(u_z^2+c_s^2 \right), \nonumber \\
k_{16}^{\star} &=& \rho \left(u_x^2+c_s^2   \right)  \left(u_y^2+c_s^2   \right), \nonumber \\
k_{17}^{\star} &=& \rho \left(u_x^2+c_s^2   \right)  \left(u_z^2+c_s^2   \right), \nonumber \\
k_{18}^{\star} &=& \rho \left(u_y^2+c_s^2   \right)  \left(u_z^2+c_s^2   \right),
\end{eqnarray}
with
\begin{eqnarray}\label{PreCollCMS}
k_5 &=& \sum_i f_i (c_{ix}^2- c_{iy}^2), \nonumber \\
k_6 &=& \sum_i f_i (c_{iy}^2- c_{iz}^2), \nonumber \\
k_7 &=& \sum_i f_i c_{ix} c_{iy}, \nonumber \\
k_8 &=& \sum_i f_i c_{ix} c_{iz}, \nonumber \\
k_9 &=& \sum_i f_i c_{iy} c_{iz}.
\end{eqnarray}
{\color{black}One can immediately observe that the RM implementation simply reduces at posing $\ket{r_i^{\star} }= \ket{k_i^{\star}}$, because $\mathbf{N} = \mathbf{I}$ in Eq.~(\ref{EqGetRawFromCentral}). A comparison between raw and central moments in terms of involved CPU time is given in Appendix~\ref{app:Q27}.}
\begin{widetext}

\section{D3Q27-CM-LBM \label{app:Q27}}
The D3Q27-CM-LBM is built by using the following lattice directions:
\begin{align}
| c_{ix}\rangle &= [0, -1,  \phantom{-}0,  \phantom{-}0, -1, -1, -1, -1,  \phantom{-}0,  \phantom{-}0, -1, -1, -1, -1,\phantom{-}1, \phantom{-}0, \phantom{-}0, \phantom{-}1,  \phantom{-}1, \phantom{-}1,  \phantom{-}1, \phantom{-}0,  \phantom{-}0, \phantom{-}1, \phantom{-}1,  \phantom{-}1, \phantom{-}1]^{\top}, \nonumber\\
| c_{iy}\rangle &= [0,  \phantom{-}0, -1, \phantom{-}0, -1, \phantom{-}1, \phantom{-}0, \phantom{-}0, -1, -1, -1, -1, \phantom{-}1, \phantom{-}1,\phantom{-}0,\phantom{-}1,\phantom{-}0,\phantom{-}1, -1,\phantom{-}0, \phantom{-}0,\phantom{-}1, \phantom{-}1,\phantom{-}1,\phantom{-}1, -1, -1]^{\top}, \nonumber\\
| c_{iz}\rangle &= [0, \phantom{-}0, \phantom{-}0, -1, \phantom{-}0, \phantom{-}0, -1, \phantom{-}1, -1, \phantom{-}1, -1,\phantom{-}1 , -1, \phantom{-}1,\phantom{-}0,\phantom{-}0,\phantom{-}1,\phantom{-}0, \phantom{-}0,\phantom{-}1, -1,\phantom{-}1, -1,\phantom{-}1, -1,\phantom{-}1, -1]^{\top}, 
\end{align}
with $i \in [0,\ldots, 26]$. The choice of these directions stems from the need to adopt the swap technique in~\citep{latt2007technical}. Following \cite{DEROSIS_PoF_31_2019}, the transformation matrix is
\begin{equation}
{\bf T} = {\bf NM} = 
\left [
	\begin{array}{c}
		\bra{|{\bf c}_i|^0} \\	
		\bra{\bar{c}_{ix}} \\	
		\bra{\bar{c}_{iy}} \\	
		\bra{\bar{c}_{iz}} \\	
		\bra{\bar{c}_{ix} \bar{c}_{iy}} \\	
		\bra{ \bar{c}_{ix} \bar{c}_{iz} } \\ 
		\bra{ \bar{c}_{iy} \bar{c}_{iz} } \\ 
		\bra{ \bar{c}_{ix}^2 - \bar{c}_{iy}^2 } \\ 
		\bra{ \bar{c}_{ix}^2 - \bar{c}_{iz}^2 } \\ 
		\bra{ \bar{c}_{ix}^2 + \bar{c}_{iy}^2 + \bar{c}_{iz}^2 } \\ 
		\bra{ \bar{c}_{ix}\bar{c}_{iy}^2 + \bar{c}_{ix}\bar{c}_{iz}^2 } \\ 
		\bra{ \bar{c}_{ix}^2\bar{c}_{iy} + \bar{c}_{iy}\bar{c}_{iz}^2 } \\ 
		\bra{ \bar{c}_{ix}^2\bar{c}_{iz} + \bar{c}_{iy}^2\bar{c}_{iz} } \\ 
		\bra{ \bar{c}_{ix}\bar{c}_{iy}^2 - \bar{c}_{ix}\bar{c}_{iz}^2 } \\ 
		\bra{ \bar{c}_{ix}^2\bar{c}_{iy} - \bar{c}_{iy}\bar{c}_{iz}^2 } \\ 
		\bra{ \bar{c}_{ix}^2\bar{c}_{iz} - \bar{c}_{iy}^2\bar{c}_{iz} } \\ 
		\bra{ \bar{c}_{ix}\bar{c}_{iy}\bar{c}_{iz} } \\ 
		\bra{ \bar{c}_{ix}^2\bar{c}_{iy}^2 + \bar{c}_{ix}^2\bar{c}_{iz}^2 + \bar{c}_{iy}^2\bar{c}_{iz}^2 } \\ 
		\bra{ \bar{c}_{ix}^2\bar{c}_{iy}^2 + \bar{c}_{ix}^2\bar{c}_{iz}^2 - \bar{c}_{iy}^2\bar{c}_{iz}^2 } \\ 
		\bra{ \bar{c}_{ix}^2\bar{c}_{iy}^2 - \bar{c}_{ix}^2\bar{c}_{iz}^2 } \\ 
		\bra{ \bar{c}_{ix}^2\bar{c}_{iy}\bar{c}_{iz} } \\ 
		\bra{ \bar{c}_{ix}\bar{c}_{iy}^2\bar{c}_{iz} } \\ 
		\bra{ \bar{c}_{ix}\bar{c}_{iy}\bar{c}_{iz}^2 } \\ 
		\bra{ \bar{c}_{ix}\bar{c}_{iy}^2\bar{c}_{iz}^2 } \\ 
		\bra{ \bar{c}_{ix}^2\bar{c}_{iy}\bar{c}_{iz}^2 } \\ 
		\bra{ \bar{c}_{ix}^2\bar{c}_{iy}^2\bar{c}_{iz} } \\ 
		\bra{ \bar{c}_{ix}^2\bar{c}_{iy}^2\bar{c}_{iz}^2 } 
	\end{array}
	\right] , \quad\quad\quad	
	{\bf M} = 
\left [
	\begin{array}{c}
		\bra{|{\bf c}_i|^0} \\	
		\bra{c_{ix}} \\	
		\bra{c_{iy}} \\	
		\bra{c_{iz}} \\	
		\bra{c_{ix} c_{iy}} \\	
		\bra{ c_{ix} c_{iz} } \\ 
		\bra{ c_{iy} c_{iz} } \\ 
		\bra{ c_{ix}^2 - c_{iy}^2 } \\ 
		\bra{ c_{ix}^2 - c_{iz}^2 } \\ 
		\bra{ c_{ix}^2 + c_{iy}^2 + c_{iz}^2 } \\ 
		\bra{ c_{ix}c_{iy}^2 + c_{ix}c_{iz}^2 } \\ 
		\bra{ c_{ix}^2c_{iy} + c_{iy}c_{iz}^2 } \\ 
		\bra{ c_{ix}^2c_{iz} + c_{iy}^2c_{iz} } \\ 
		\bra{ c_{ix}c_{iy}^2 - c_{ix}c_{iz}^2 } \\ 
		\bra{ c_{ix}^2c_{iy} - c_{iy}c_{iz}^2 } \\ 
		\bra{ c_{ix}^2c_{iz} - c_{iy}^2c_{iz} } \\ 
		\bra{ c_{ix}c_{iy}c_{iz} } \\ 
		\bra{ c_{ix}^2c_{iy}^2 + c_{ix}^2c_{iz}^2 + c_{iy}^2c_{iz}^2 } \\ 
		\bra{ c_{ix}^2c_{iy}^2 + c_{ix}^2c_{iz}^2 - c_{iy}^2c_{iz}^2 } \\ 
		\bra{ c_{ix}^2c_{iy}^2 - c_{ix}^2c_{iz}^2 } \\ 
		\bra{ c_{ix}^2c_{iy}c_{iz} } \\ 
		\bra{ c_{ix}c_{iy}^2c_{iz} } \\ 
		\bra{ c_{ix}c_{iy}c_{iz}^2 } \\ 
		\bra{ c_{ix}c_{iy}^2c_{iz}^2 } \\ 
		\bra{ c_{ix}^2c_{iy}c_{iz}^2 } \\ 
		\bra{ c_{ix}^2c_{iy}^2c_{iz} } \\ 
		\bra{ c_{ix}^2c_{iy}^2c_{iz}^2 } 
	\end{array}
	\right] ,
\end{equation}
where $\mathbf{N} = \mathbf{T}\mathbf{M}^{-1}$. Post-collision populations are computed by the ``two-step" reconstruction~\citep{Fei2017}. Indeed, we first compute post-collision raw moments $\ket{r_i^{\star}} = \left[ r_0^{\star},\, \ldots, \, r_i^{\star},\, \ldots, \, r_{26}^{\star}   \right]^{\top}$ as
\begin{equation}
\ket{r_i^{\star}} 					  =  {\mathbf{N}^{-1}} \ket{k_i^{\star} },
\end{equation}
where $k_i^{\star}$ are given in~\citep{DEROSIS_PoF_31_2019}. The expressions of $r_i^{\star}$ are too tedious and are not reported in the following.
The interested reader can refer to the script D3Q27CentralMoments.m in the supplementary material in order to derive $\ket{r_i^{\star}} $. Then, we transform into populations
\begin{equation}
\ket{f_i^{\star}}  = \mathbf{M}^{-1} \ket{r_i^{\star}},
\end{equation}
i.e.
\begin{eqnarray} \label{pdfQ27}
f_{0}^{\star} &=&  r_0^{\star} - r_9^{\star} + r_{17}^{\star} - r_{26}^{\star},\nonumber \\
f_{1}^{\star} &=&   (r_7^{\star} + r_8^{\star} + r_9^{\star})/6 - (r_{17}^{\star} + r_{18}^{\star})/4 +(r_{10}^{\star} + r_{26}^{\star}  - r_1^{\star} - r_{23}^{\star})/2 ,\nonumber \\
f_{2}^{\star} &=&  (r_{18}^{\star} - 3 r_{17}^{\star})/8 + (r_8^{\star} + r_9^{\star})/6  - r_{19}^{\star}/4 - r_7^{\star}/3   + (r_{11}^{\star} + r_{26}^{\star}- r_2^{\star} - r_{24}^{\star})/2,\nonumber \\
f_{3}^{\star} &=&  (r_{18}^{\star} - 3 r_{17}^{\star})/8 + (r_7^{\star} + r_9^{\star})/6+ r_{19}^{\star}/4 - r_8^{\star}/3    + (r_{12}^{\star} + r_{26}^{\star} - r_3^{\star} - r_{25}^{\star})/2,\nonumber \\
f_{4}^{\star} &=& ( r_{17}^{\star} + r_{18}^{\star})/16 + (r_{19}^{\star}- r_{10}^{\star} - r_{11}^{\star} - r_{13}^{\star} - r_{14}^{\star})/8  + (r_4^{\star} + r_{23}^{\star} + r_{24}^{\star} - r_{22}^{\star} - r_{26}^{\star})/4,\nonumber \\
f_{5}^{\star} &=&  (r_{17}^{\star} + r_{18}^{\star})/16 + (r_{11}^{\star} + r_{14}^{\star} + r_{19}^{\star}  - r_{10}^{\star} - r_{13}^{\star})/8  + (r_{22}^{\star} + r_{23}^{\star}- r_4^{\star} - r_{24}^{\star} - r_{26}^{\star})/4,\nonumber \\
f_{6}^{\star} &=&  (r_{17}^{\star} + r_{18}^{\star})/16 + (r_{13}^{\star} - r_{10}^{\star} - r_{12}^{\star} - r_{15}^{\star} - r_{19}^{\star})/8 + (r_5^{\star} + r_{23}^{\star} + r_{25}^{\star} - r_{21}^{\star} - r_{26}^{\star})/4,\nonumber \\
f_{7}^{\star} &=& ( r_{17}^{\star} + r_{18}^{\star})/16 + (r_{12}^{\star} + r_{13}^{\star} + r_{15}^{\star} - r_{10}^{\star} - r_{19}^{\star})/8 + (r_{21}^{\star} + r_{23}^{\star}  - r_5^{\star} - r_{25}^{\star} - r_{26}^{\star})/4,\nonumber \\
f_{8}^{\star} &=&  (r_{14}^{\star} + r_{15}^{\star} + r_{17}^{\star} - r_{11}^{\star} - r_{12}^{\star} - r_{18}^{\star})/8  + (r_6^{\star} + r_{24}^{\star} + r_{25}^{\star} -  r_{20}^{\star} - r_{26}^{\star})/4,\nonumber \\
f_{9}^{\star} &=&  (r_{12}^{\star} + r_{14}^{\star} + r_{17}^{\star} - r_{11}^{\star} - r_{15}^{\star} - r_{18}^{\star})/8 + (r_{20}^{\star} + r_{24}^{\star} - r_6^{\star} - r_{25}^{\star} - r_{26}^{\star})/4 ,\nonumber \\
f_{10}^{\star} &=&  (r_{20}^{\star} + r_{21}^{\star} + r_{22}^{\star} + r_{26}^{\star} - r_{16}^{\star} - r_{23}^{\star} - r_{24}^{\star} - r_{25}^{\star})/8,\nonumber \\
f_{11}^{\star} &=&  (r_{16}^{\star} + r_{22}^{\star} + r_{25}^{\star} + r_{26}^{\star} - r_{20}^{\star} - r_{21}^{\star} - r_{23}^{\star} - r_{24}^{\star})/8,\nonumber \\
f_{12}^{\star} &=&  (r_{16}^{\star} + r_{21}^{\star} + r_{24}^{\star} + r_{26}^{\star} - r_{20}^{\star} - r_{22}^{\star} - r_{23}^{\star} - r_{25}^{\star})/8,\nonumber \\
f_{13}^{\star} &=&  (r_{20}^{\star} + r_{24}^{\star} + r_{25}^{\star} + r_{26}^{\star} - r_{16}^{\star} - r_{21}^{\star} - r_{22}^{\star} - r_{23}^{\star})/8,\nonumber \\
f_{14}^{\star} &=&  (r_7^{\star} + r_8^{\star} + r_9^{\star})/6 - (r_{17}^{\star} + r_{18}^{\star})/4 + (r_1^{\star} + r_{23}^{\star} + r_{26}^{\star}-r_{10}^{\star})/2,\nonumber \\
f_{15}^{\star} &=&  (r_{18}^{\star}  - 3 r_{17}^{\star})/8 + (r_8^{\star} + r_9^{\star})/6 - r_{19}^{\star}/4 - r_7^{\star}/3   + (r_2^{\star} + r_{24}^{\star} + r_{26}^{\star}- r_{11}^{\star})/2,\nonumber \\
f_{16}^{\star} &=&  (r_{18}^{\star} - 3 r_{17}^{\star})/8 + (r_7^{\star} + r_9^{\star})/6 + r_{19}^{\star}/4 - r_8^{\star}/3  + (r_3^{\star} + r_{25}^{\star} + r_{26}^{\star}- r_{12}^{\star})/2,\nonumber \\
f_{17}^{\star} &=&  (r_{17}^{\star} + r_{18}^{\star})/16 + (r_{10}^{\star} + r_{11}^{\star} + r_{13}^{\star} + r_{14}^{\star} + r_{19}^{\star})/8 + (r_4^{\star}/4 - r_{22}^{\star} - r_{23}^{\star} - r_{24}^{\star} - r_{26}^{\star})/4 ,\nonumber \\
f_{18}^{\star} &=&  (r_{17}^{\star} + r_{18}^{\star})/16 + (r_{10}^{\star} + r_{13}^{\star} + r_{19}^{\star} - r_{11}^{\star} - r_{14}^{\star})/8  + (r_{22}^{\star} + r_{24}^{\star}- r_4^{\star} - r_{23}^{\star} - r_{26}^{\star})/4,\nonumber \\
f_{19}^{\star} &=&  (r_{17}^{\star} + r_{18}^{\star})/16 + (r_{10}^{\star} + r_{12}^{\star} + r_{15}^{\star} - r_{13}^{\star} - r_{19}^{\star})/8 + (r_5^{\star}/4- r_{21}^{\star} - r_{23}^{\star} - r_{25}^{\star} - r_{26}^{\star})/4  ,\nonumber \\
f_{20}^{\star} &=&  (r_{17}^{\star} + r_{18}^{\star})/16 + (r_{10}^{\star} - r_{12}^{\star} + r_{13}^{\star} + r_{15}^{\star} + r_{19}^{\star})/8+ (r_{21}^{\star} + r_{25}^{\star} - r_5^{\star} - r_{23}^{\star} - r_{26}^{\star})/4 ,\nonumber \\
f_{21}^{\star} &=&  (r_6^{\star} - r_{20}^{\star} - r_{24}^{\star} - r_{25}^{\star} - r_{26}^{\star})/4 + (r_{11}^{\star} + r_{12}^{\star} + r_{17}^{\star} - r_{14}^{\star} - r_{15}^{\star} - r_{18}^{\star})/8,\nonumber \\
f_{22}^{\star} &=&  (r_{20}^{\star} + r_{25}^{\star} - r_6^{\star} - r_{24}^{\star} - r_{26}^{\star})/4 + (r_{11}^{\star} + r_{15}^{\star} + r_{17}^{\star} - r_{12}^{\star} - r_{14}^{\star} - r_{18}^{\star})/8 ,\nonumber \\
f_{23}^{\star} &=&  (r_{16}^{\star} + r_{20}^{\star} + r_{21}^{\star} + r_{22}^{\star} + r_{23}^{\star} + r_{24}^{\star} + r_{25}^{\star} + r_{26}^{\star})/8,\nonumber \\
f_{24}^{\star} &=&   (r_{22}^{\star} + r_{23}^{\star} + r_{24}^{\star} + r_{26}^{\star} - r_{16}^{\star} - r_{20}^{\star} - r_{21}^{\star} - r_{25}^{\star})/8,\nonumber \\
f_{25}^{\star} &=&  (r_{21}^{\star} + r_{23}^{\star} + r_{25}^{\star} + r_{26}^{\star} - r_{16}^{\star} - r_{20}^{\star} - r_{22}^{\star} - r_{24}^{\star})/8,\nonumber \\
f_{26}^{\star} &=&  (r_{16}^{\star} + r_{20}^{\star} + r_{23}^{\star} + r_{26}^{\star} - r_{21}^{\star} - r_{22}^{\star} - r_{24}^{\star} - r_{25}^{\star})/8.
\end{eqnarray}
\end{widetext}

{\color{black}Eventually, we use the test case in Section~\ref{3DTG} to compare the CPU time required by the D3Q19 and D3Q27 lattice discretizations, as well as, the collision models (raw and central moments). 
In addition, we carry out different simulations by varying the number of grid points in each direction ($D=32$, $48$, $64$, $96$ and $128$), and the corresponding CPU time is recorded for each run.  All measurements are performed on a iMac 27 equipped with Intel Core i5 6-core 3.3GHz and 8GB of RAM. Before moving to the results, it is worth noting that in the present context, collision models are based on the equilibration of (1) bulk-viscosity-related and (2) higher-order moments, which allows for a non-negligible reduction of the CPU time as compared to their full MRT formulation. This will be discussed in more details in another study.

In Figure~\ref{timing}, the normalized CPU time is plotted as a function of $D^3$. Globally speaking, all configurations show that the involved computational time grow with the fourth power of the total number of points. 
This is in accordance with the fact that the CPU time is proportional to $D^3 \times T_{ite}$, where the number of time iterations $T_{ite}$ linearly depends on $D$ when the time step is computed via an acoustic scaling (i.e., $T_{ite} \propto D$). 
Quantitavely speaking, the adoption of CMs increases the CPU time of $\sim 7\%$ for the D3Q19-LBM, as compared to its RMs counterpart (see Table~\ref{TableTiming}). 
Moving to 27 discrete velocities, this gap further increases between the RM and CM formulations, and it reaches $\sim 20\%$.
Moreover, we find that the D3Q19-CM-LBM allows us to save a considerable amount of computational time ($\sim 70 \%$), as compared to the more general D3Q27 formulation.
\begin{figure}[!htbp]
\centering
\includegraphics[scale=1]{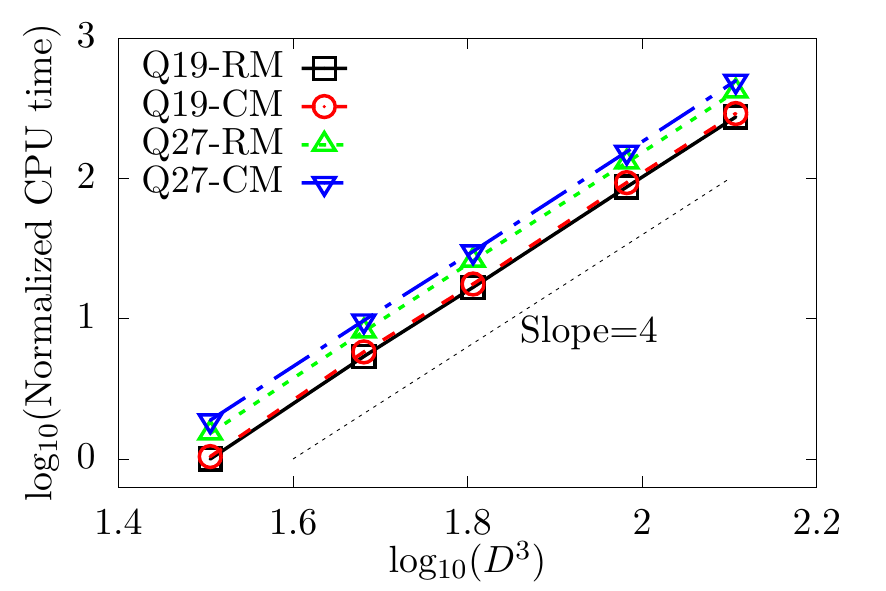}
\caption{Computational cost: normalized CPU time measured for the D3Q19-RM (black solid line and squares), the D3Q19-CM (red dashed line and circles), the D3Q27-RM (green dotted line and triangles), and the D3Q27-CM (blue dash-dotted line and inverted triangles). The black dotted line has slope equal to 4.}
\label{timing}
\end{figure}
\begin{table}[htbp!]
\centering
\begin{tabular}{ c | c | c | c  | c }
\hline
\hline
$D$ & Q19-RM & Q19-CM & Q27-RM & Q27-CM\\
\hline
32 & 1 & 1.04  & 1.53 & 1.89\\
48  & 5.40 & 5.81 & 8.14 & 9.74\\
64 & 16.78 & 17.74 & 25.94 & 30.24\\
96 & 87.11 & 93.58 & 130.41 & 155.47\\
128 & 274.00 & 290.99 & 417.8545 & 498.12\\
\hline
\hline
\end{tabular}
\caption{Computational cost: normalized CPU time measured for two lattices (D3Q19 and D3Q27), as well as, two collision models (RMs and CMs). In the present context, higher-order and bulk-viscosity-reltated moments are equilibrated for all configurations.}
\label{TableTiming}
\end{table}

}

\section{D3Q19-CM-LBM color-gradient \label{app:AppCG}}

{\color{black}The color-gradient method (CGM) was originally proposed by Rothman \& Keller~\cite{rothman1988immiscible}, but their model was not able to account for density and contrasts. This feature was developed by Grunau \textit{et al.}~\cite{grunau1993lattice} in 1993 by means of a hexagonal lattice and later by Reis \& Phillips~\cite{Reis_2007} with the D2Q9 discretization. Then, Leclaire \textit{et al.}~\cite{LECLAIRE20122237} et al. adapted the recoloring operator in Ref.~\cite{PhysRevE.71.056702} for the model in Ref.~\cite{Reis_2007} in the case of variable density ratios. It should be note that the original Rothman \& Keller CGM has been successfully employed to simulate flows in heterogeneous porous media \cite{bakhshian2020new,bakhshian2020scaling}. Hereafter, we will recall the basic features of the D3Q19 formulation of the CGM.}

Let us consider two immiscible fluids, namely, red and blue. The evolution of populations $f_i^k$ is
\begin{equation}
f_i^k(\bm{x}+\bm{c}_i, t+1) = f_i^k(\bm{x}, t) + \Omega_i^k(\bm{x}, t),
\end{equation} 
where $k=r$ for the red fluid, and $k=b$ for the blue one. Moreover, it is possible to define the total distribution functions as $\displaystyle f_i = f_i^r + f_i^b$. The collision process $\Omega_i^k$ is composed of three sub-stages:
\begin{equation}
\Omega_i^k = \big(\Omega_i^k \big)^{(3)} \left[  \left(  \Omega_i^k \right)^{(1)} + \left(  \Omega_i^k \right)^{(2)}   \right],
\end{equation}
where $ \left( \Omega_i^k \right)^{(1)}$, $ \left( \Omega_i^k \right)^{(2)}$ and $ \left( \Omega_i^k \right)^{(3)}$ are the single-phase, perturbation and recoloring operators, respectively. Macroscopic variables are given by
\begin{eqnarray}
\rho_k &=& \sum_i f_i^k, \nonumber \\
\rho &=& \sum_k \rho_k,  \nonumber \\
\rho \bm{u} &=& \sum_i f_i  \bm{c}_i + \frac{1}{2}\bm{F},
\end{eqnarray}
where $\rho_k$ is the density of the fluid $k$, $\rho$ is the total mass density, $\bm{u}$ is the total momentum and $\bm{F}$ is a body force. The single-phase collision operator is
\begin{equation}
\left( \Omega_i^k \right)^{(1)} = \mathbf{T}^{-1} \mathbf{K}  \mathbf{T} \left(\ket{h_i^{\mathrm{eq}}} - \ket{f_i} \right) + \mathbf{T}^{-1} \left( \mathbf{I} - \frac{\mathbf{K}}{2} \right)  \ket{R_i},
\end{equation}
where $h^{eq}_i$ is the equilibrium used by the color-gradient method:
\begin{equation}
h^{eq}_i = f^{eq}_i + \rho(\phi_i-w_i)\label{eq:feqCG}
\end{equation}
with $f^{eq}_i$ the standard version of the equilibrium~(\ref{eq:EqQ19RM}), and
\begin{equation}
	\phi_i = 
	\begin{cases}
		~\alpha, & |\bm{c}_i|^2=0,\\
		~(1-\alpha)/12, & |\bm{c}_i|^2=1, \\
		~(1-\alpha)/24, & |\bm{c}_i|^2=2.
		\label{eq:varphi_ik}
\end{cases}
\end{equation}
Notice that $\phi=0$ for the original single phase collision operator. To distinguish the two components, the order parameter $\phi$ is introduced, that is
\begin{equation}
	\phi = \frac{\rho_r-\rho_b}{\rho_r+\rho_b}.
	\label{eq:order}
\end{equation}
The values $\phi=1,-1$, and $0$ correspond to a purely red fluid, a purely blue fluid, and the interface, respectively. To obtain a stable interface, the density ratio between the fluids must be taken into account as follows to obtain a stable interface~\cite{grunau1993lattice}:
\begin{equation}
\frac{\rho_r^0}{\rho_b^0} = \frac{1-\alpha_b}{1-\alpha_r},
\end{equation}
where the superscript ``0'' indicates the initial value of the density at the beginning of the simulation~\cite{PhysRevE.95.033306}. The pressure of the fluid is given as an isothermal equation of state:
\begin{equation}
p = \rho \left(c_s^k \right)^2 = \frac{1}{2}\rho_k (1-\alpha),
\label{eq:press}
\end{equation}
for the D3Q19 lattice, where $c_s^k$ is the speed of sound of the fluid $k$~\cite{wen2019improved}, $\alpha$ is interpolated by
\begin{equation}
\alpha = \frac{\rho_r \alpha_r + \rho_b \alpha_b }{\rho_r  + \rho_b},
\end{equation}
with $\alpha_b = 1/3$ and $c_s^b=1/\sqrt{3}$~\cite{saito2018color}.\\
\indent Following~\cite{BRACKBILL1992335, Reis_2007, PhysRevE.85.046309, wen2019improved}, the interfacial tension is modeled by the so-called perturbation operator:
\begin{equation}
	\left(\Omega_i \right)^{(2)} = \frac{A_k}{2} |\nabla \phi| 
	\left[w_i \frac{(\bm{c}_i \cdot \nabla \phi)}{|\nabla \phi|^2} - B_i \right],
	\label{eq:perturb}
\end{equation}
where
\begin{equation}
	B_i = 
	\begin{cases}
		~-1/3, & |\bm{c}_i|^2=0,\\
		~+1/18, & |\bm{c}_i|^2=1, \\
		~+1/36, & |\bm{c}_i|^2=2. \label{eq:B_i}
\end{cases}
\end{equation} 
and coefficients $A_k$ are related to the surface tension $\sigma$ as
\begin{equation}
\sigma = \frac{2}{9}\frac{A_b+A_r}{\omega}, \qquad \mathrm{with} \quad A_r=A_k.
\end{equation}
\indent Eventually, the following recoloring operator is applied to promote phase segregation and maintain the interface:
\begin{eqnarray}
	(\Omega_i^r)^{(3)} = \frac{\rho_r}{\rho}f_i 
	+ \beta \frac{\rho_r \rho_b}{\rho^2} \cos(\theta_i)
	g_i^{eq}(\rho,\bf{0}), \label{eq:op3r} \nonumber
\\
	(\Omega_i^b)^{(3)} = \frac{\rho_b}{\rho}f_i 
	- \beta \frac{\rho_r \rho_b}{\rho^2} \cos(\theta_i)
	g_i^{eq}(\rho,\bf{0}), \label{eq:op3b}
\end{eqnarray}
where $\beta=0.7$~\cite{LECLAIRE20122237, saito2017lattice, saito2018color}, $\theta_i$ is the angle between $\nabla\phi$ and $\bm{c}_i$, which is defined by
\begin{equation}
	\cos(\theta_i) = \frac{\bm{c}_i \cdot \nabla \phi}{|\bm{c}_i|  |\nabla \phi|}.
\end{equation}

Notably, the post-collision state in terms of populations and raw moments maintain the same form shown in App.~\ref{app:Q27}. The only change in the collision process affects the post-collision CMs that are reported in the following for the sake of completeness:
\begin{eqnarray}
k_0^{\star} &=& \rho, \nonumber \\
k_1^{\star} &=& F_x/2, \nonumber \\
k_2^{\star} &=& F_y/2, \nonumber \\
k_3^{\star} &=& F_z/2, \nonumber \\
k_4^{\star} &=& 3 \rho \left(1- \alpha \right)/2, \nonumber \\
k_5^{\star} &=& \left(1-\omega \right)k_5, \nonumber \\
k_6^{\star} &=& \left(1-\omega \right)k_6 \nonumber \\
k_7^{\star} &=& \left(1-\omega \right)k_7, \nonumber \\
k_8^{\star} &=& \left(1-\omega \right)k_8, \nonumber \\
k_9^{\star} &=& \left(1-\omega \right)k_9, \nonumber \\
k_{10}^{\star} &=& F_y c_s^2 /2 + \rho u_y \left( 3 \alpha-1 \right)/6, \nonumber \\
k_{11}^{\star} &=& F_x c_s^2 /2 + \rho u_x \left( 3 \alpha-1 \right)/6, \nonumber \\
k_{12}^{\star} &=& F_z c_s^2 /2 + \rho u_z \left( 3 \alpha-1 \right)/6, \nonumber \\
k_{13}^{\star} &=& F_x c_s^2 /2 + \rho u_x \left( 3 \alpha-1 \right)/6, \nonumber \\
k_{14}^{\star} &=& F_z c_s^2 /2 + \rho u_z \left( 3 \alpha-1 \right)/6, \nonumber \\
k_{15}^{\star} &=& F_y c_s^2 /2 + \rho u_y \left( 3 \alpha-1 \right)/6, \nonumber \\
k_{16}^{\star} &=& \rho \left[ 1 + u_x^2 + u_y^2 - \alpha \left(1+3u_x^2+3u_y^2 \right)  \right], \nonumber \\
k_{17}^{\star} &=& \rho \left[ 1 + u_x^2 + u_z^2 - \alpha \left(1+3u_x^2+3u_z^2 \right)  \right], \nonumber \\
k_{18}^{\star} &=& \rho \left[ 1 + u_y^2 + u_z^2 - \alpha \left(1+3u_y^2+3u_z^2 \right)  \right].
\end{eqnarray} 
\bibliography{bibliography.bib}
\end{document}